\documentstyle[12pt,a4wide,amsmath,amssymb,epsf]{article} 

\newcommand{\nin}{\noindent}
\newcommand{\be}{\begin{equation}}
\newcommand{\ee}{\end{equation}}
\newcommand{\bea}{\begin{eqnarray}}
\newcommand{\eea}{\end{eqnarray}}
\newcommand{\br}{\hskip .25cm/\hskip -.25cm}
\newcommand{\hf}{\frac{1}{2}}
\newcommand{\nonu}{\nonumber\\}

\newcommand{\dg}{^\dagger}
\newcommand{\ol}{\overline}

\begin{document}

\begin{titlepage} 

\begin{center}

{\Large \textbf{Towards Exact Results in Nodal\\
Antiferromagnetic Planar Liquids}}

\vspace{1cm}

{\bf J.~Alexandre}, {\bf N.E.~Mavromatos} and {\bf Sarben Sarkar}\\[0pt]
Department of Physics, Theoretical Physics, King's College London\\[0pt]
Strand, London WC2R 2LS, UK

\vspace{2cm}

\textbf{Abstract}

\end{center}

\noindent {\small It has been argued in previous works 
by the authors that nodal excitations in (2+1)-dimensional 
doped antiferromagnets might
exhibit, in the spin-charge separation framework and at 
specific regions of the parameter space, a supersymmetry 
between spinons and holons. This 
supersymmetry has been elevated to a $N=2$
extended supersymmetry of composite operators of spinon and holons,
corresponding to the effective ``hadronic" degrees of freedom.
In this work we elaborate further on this idea by describing 
in some detail the dynamics of a
specific composite model corresponding to an {\it Abelian}
Higgs model (SQED). The abelian nature of the gauge group 
seems to be necessitated both 
by the composite structure used, but also by electric charge 
considerations for the various composites. We demonstrate 
the passage from a pseudogap to an unconventional superconducting phase,
which notably  
is an exact non-perturbative analytic result, due to the underlying 
$N=2$ supersymmetric Abelian gauge theory. 
We believe that these considerations 
may provide a first step towards a non-perturbative understanding 
of the phase diagrams of strongly-correlated electron systems.}

\end{titlepage}

\section{Introduction and Overview of Results} 

\bigskip The study of $2+1$ dimensional gauge theories has been motivated by
the search for an understanding of confinement \cite{conf}. The degree to
which they can be of help in understanding confinement at zero temperature
in three space dimensions is unclear because the nature and possible
configurations of textures, instantons and other non-perturbative features
such as the locality of disorder variables is very different. Their
attraction is that their comparative simplicity allows a much more complete
understanding of the confinement phenomenon. However another reason for
studying such theories is the interest in planar system in high temperature
superconductors \cite{var} where the phases are believed not to be of
conventional type. In particular string-like 
structures which are features
of confinement in the gauge theories occur as inhomogeneities in the charge
and magnetic order \cite{stripe} in certain parameter regimes of these
materials. Both in the study of confinement and the condensed matter
analogues, some of the reason for controversy is the room for error in the
calculations that can be performed due to the intrinsic limitations of the
methods themselves. It is thus important to pursue models with exact
solutions. Unlike for $1+1$ dimensions where infinite-dimensional groups and
the factorizability of the S-matrix are at play \cite{Zam},\ dynamical
supersymmetry is an important ingredient for obtaining exact information
about the phases of the theory \cite{seiberg,intril,deboer,strassler}. 
In fact, the more extended the supersymmetry the more exact the results.
This is related to holomorphicity properties of the theory. 

Speaking about supersymmetry in condensed matter systems may at first sight
seem absurd. After all, supersymmetry is a space time symmetry, requiring
Lorentz invariance, which condensed matter systems do not have, either 
because they are primarily lattice systems, or, in case one considers
effective {\it continuous} quantum field theories of such systems, 
because there are finite fermi surfaces. Excitations near finite 
fermi surfaces cannot produce relativistic (Lorentz invariant)
field theories. 
However, there are cases where relativistic low-energy effective
quantum field theories do appear in condensed matter situations. 
One famous example is the $CP^1$ $\sigma$-model describing the 
low-energy limit of an antiferromagnet. Other examples, in which we shall
concentrate in this article, include excitations near {\it nodal} points
in the fermi surface, i.e. near points where either the fermi surface
shrinks to a set of points, as is the case of underdoped cuprates, 
or a gap function 
vanishes (as is the case of $d$-wave high-temperature superconductors).

It is our view, which is also shared by a large proportion of the 
community, that such nodes play an important r\^ole
on the rich phase diagram of these materials. 
It is therefore of outmost importance to look carefully at the 
physics implied by excitations near such nodes, which, from 
a continuum effective field theory viewpoint,  constitute
a {\it relativistic} field theoretic system, in which the 
role of the limiting velocity (`speed of light') is played
by the fermi velocity of the node. 

One can go one step further:  
upon assuming separation of spin and electric charge 
degrees of freedom in the fundamental constituents describing the 
ground state of {\it planar} doped antiferromagnets (or more generally strongly
correlated electronic systems), one observes that there is a balance
between fermionic and bosonic degrees of freedom~\footnote{In (2+1)-dimensions
the concept of a boson or fermion is not well defined, and one has 
the possibility of fractional statistics as well. Formally this can
be easily understood by means of bosonization techniques~\cite{marchetti}.
However, the concept of bosonic (commuting) or fermionic (anticommuting)
operator variables can be defined, and it is in this sense that 
the we use the terminology fermion and boson.}. This balance prompts 
one to seek for possible supersymmetries in certain regions of the 
parameter space of the underlying condensed matter system.

In \cite{diamand,mav} it was argued that it is indeed possible to 
determine regions in the parameter space of 
condensed matter models, of relevance to 
the physics of doped antiferromagnets (and hence high-temperature
superconductors), where there is such a 
{\it dynamical supersymmetry} between appropriate degrees of freedom.
Specifically, 
one starts from a microscopic lattice system,
an appropriately extended $t-j$ model~\cite{mav} with nodes in its 
fermi surface.  It is an experimental fact that nodes exist
in {\it both} the underdoped regime of the cuprates (where the fermi surface
consists of four nodes), and in the superconducting regime
(the high-$T_c$ oxides are $d$-wave superconductors, with nodes in 
the respective superconducting gap). 
The 
continuum 
low-energy theory of {\it nodal} spinons and holon excitations 
has the form of a relativistic   
$CP^1$ $\sigma$-model (magnon-spinons) coupled to Dirac-like fermions 
(holon degrees of freedom) plus higher contact interactions among the 
various field modes. 
In \cite{mav} it was shown that there are regions
in the microscopic
model phase space, where one recovers a supersymmetric 
theory between the spinon and holon constituents in the presence of 
strongly-coupled non-dynamical (i.e. no kinetic terms)  
gauge fields. 
This is the constituent theory. 
The non-dynamical gauge fields of their $CP^1$ $\sigma$-model simply express
contact interactions between spinons and holons.

Although $N=1$ symmetry was
demonstrated for specific condensed matter models of doped 
antiferromagnets~\cite{diamand,mav}, however a lack of 
holomorphicity
properties meant that exact non-perturbative 
information was not available.
The situation is much improved for $N=2$
supersymmetry~\cite{intril,strassler}, and such an elevation
is indeed necessitated by the electric charge conservation 
requirement~\cite{diamand}. It is the point of this paper to 
construct a precise N=2 supersymmetric model of composites of 
spinons and holons, which arguably described the low energy dynamics
of nodal liquids, and discuss the associated physics, especially 
in connection with a passage from the pseudogap to the superconducting
phase.

Although at first sight the model of \cite{mav}
appears to have only a $N=1$ supersymmetry, however, as argued in
\cite{ams} it actually has a 
hidden $N=2$ supersymmetry~\cite{compos} due to its 
low dimensionality (2+1 dimensions), which implies the 
existence of topologically conserved currents (i.e. currents which are 
conserved without use of the equations of motion). This is a 
generic result for such theories~\cite{hlousek,alex}. 
In~\cite{compos,ams} 
we demonstrated how the $N=1$ supersymmetry, 
in terms of constituent
fields can be elevated to a $N=2$ supersymmetry in terms of 
suitably
constructed composite fields. 
As shown in \cite{compos,ams}, once can generate this way 
{\it dynamical gauge fields},
at the composite operator level,
obtained after integrating out the non-dynamical gauge fields
of the constituent theory. 
Such gauge structures are 
made out of appropriate combinations of spinons and holons,
up to {\it quartic order} in constituent fields. 
The presence of higher order composites, in contrast to the bilinear
composite models of the non supersymmetric case of ref.~\cite{farakos}, 
is necessary here~\cite{ams}, in order to guarantee the 
gauge coupling of the initial $N=1$ scalar and vector supermultiplets.

It is interesting to notice that the choice of composite operators 
made in \cite{compos,ams} led explicitly to 
an Abelian gauge field involved in the N=2 supersymmetric
multiplet, which essentially coupled
two $N=1$ composite supermultiplets, a scalar and a vector. 
We were unable to find a choice of composite operators
that generate the full Non-Abelian SU(2) supersymmetric model of 
\cite{affleck}, whose breaking $SU(2) \to U(1)$ would result 
in the Abelian Higgs composite model discussed explicitly 
in \cite{compos}. 
The two cases (N=2 Super QED and a Broken phase of 
N=2 SU(2) Georgi-Glashow model)
lead to very different 
non-perturbative dynamics~\cite{strassler}, for instance as far as 
confinement properties of the three-dimensional gauge theory are concerned.
In the latter case there is no stable monopole phase if supersymmetry is 
unbroken. We remind the reader that in non supersymmetric theories,
Polyakov has shown that a stable monopole-plasma phase in (2+1)-dimensions
lad to a massive photon, and a linear confinement. In the 
N=2 SU(2) case, such a phase does not exist~\cite{strassler},
and this latter result is understood to be exact.  

It is the purpose of this article to 
discuss the very different physics of the two above mentioned
{\it distinct} cases $N=2$ Supersymmetric QED (SQED) (with compact $U(1)$
group) and $N=2$ SU(2) $\to$ U(1) broken phase of a non-Abelian model
discussed in \cite{affleck}. The difference is exemplified by the
very different confinement properties. We shall demonstrate, though, 
that the non compact case also presents interesting physics, 
namely it allows an exact study of the passage of a pseudogap to a 
superconducting phase for the nodal liquid. The pseudogap phase
of the non compact case will correspond to the Higgs phase 
of the SQED, while the superconducting phase, which is characterised 
by unconventional superconductivity of the anomalous type proposed in 
\cite{dorey}, corresponds to the Coulomb-phase of the SQED, in which 
the statistical gauge field remains exactly massless. 

Our results may be of relevance in 
attempts towards an 
analytic understanding at a non-perturbative level of 
the dynamics of strongly-correlated electron systems, 
with relevance to high-temperature superconductivity and more generally
to antiferromagnetism.  
Also our model may be viewed
as a toy model for an understanding of ideas related to 
the so-called {\it scaleless limit} 
of 
gauge theories~\cite{bjorken}, where the gauge fields appear 
dynamically from more fundamental interacting constituents in the theory.

However, in the four-dimensional setting of \cite{bjorken}, the emergent  
gauge bosons (photons) appeared as Goldstone bosons of a spontaneous 
breakdown of Lorentz symmetry, associated with non-zero vacuum expectation
values (v.e.v.) 
of vector fields linearising the four-fermion 
Thirring interactions of the model. 
In three space-time dimensions, on the other hand, 
the photons have only a single degree of freedom and hence they are allowed,
in a sense, to get a v.e.v 
without breaking Lorentz symmetry. 
It is this fact that allows the extension of such ideas in (2+1)-dimensions
to incorporate supersymmetry, which is intimately related to Lorentz
invariance. From a physical point of view, with relevance to strongly-
correlated electrons, we remark that 
the maintenance of Lorentz symmetry is connected with 
the fact that we restrict our attention  to
excitations near nodes of the fermi surface of such 
systems~\cite{mav}, which exhibit relativistic behaviour, with 
the r\^ole of the velocity of `light'  played by the fermi 
velocity at the node. 

The structure of the article is as follows: in section 2 we review the 
construction (in the continuum) of the N=2 supersymmetric 
composite model. In section 3 we discuss the electric charge assignment
of the composite excitations, which is crucial in determining the 
physically correct effective action to describe the continuum
composite dynamics. In section 4 we apply these considerations to 
a specific microscopic model of strongly-correlated electrons~\cite{mav},
and discuss the emergence of the correct continuum limit. Moreover, 
in the same section we discuss exact results concerning the 
phase structure of this effective continuum 
composite theory, and demonstrate the 
passage from an (unconventional~\cite{dorey}) 
superconducting to a pseudogap (and stripe) phase.
We also discuss an exact result concerning the non-fermi liquid behaviour
of our composite nodal supersymmetric liquid. 
Section 5 contains our conclusions and outlook. Technical aspects of
our approach, which regard the formalism and the phase structure 
of N=2 (2+1)-dimensional
supersymmetric gauge theories~\cite{intril,deboer}, 
are given in an Appendix.  

\section{Review of the construction of the (continuum) 
N=2 Supersymmetric Abelian Composite Model}

We will first summarize the results of \cite{compos,ams}, dealing 
with the construction of $N=2$ 
composite supermultiplets, quartic in the 
constituent fields of holons and spinons. 

The $\gamma-$matrices are $\gamma^0=\sigma^2$, $\gamma^1=i\sigma^1$ and $\gamma^2=i\sigma^3$, 
where $\sigma^1,\sigma^2,\sigma^3$ are the Pauli matrices. We define for any spinor $\psi$:
$\ol\psi=\psi\dg\gamma^0$.

First we note that, as discussed in \cite{compos}, the quadratic composites

\begin{eqnarray}  \label{defcomp}
\phi&=&\overline\psi_1\psi_2 \\
A_\mu&=&\overline\psi_1\gamma_\mu\psi_2-z_1^\star\partial_\mu
z_2+z_2\partial_\mu z_1^\star  \nonumber
\end{eqnarray}

\noindent belong respectively to a $N=1$ scalar supermultiplet $%
(\phi',\psi',f) $ and a $N=1$ vector supermultiplet $(A_\mu,\chi')$ were the
composite superpartners read, in terms of the constituent fields

\begin{eqnarray}  \label{quadpartners}
\psi'&=&(f_1^\star-\hskip .25cm/\hskip -.25cm\partial z_1^\star)\psi_2+(f_2-%
\hskip .25cm/\hskip -.25cm\partial z_2)\psi_1^\star  \nonumber \\
\chi'&=&(f_1^\star+\hskip .25cm/\hskip -.25cm\partial z_1^\star)\psi_2-(f_2+%
\hskip .25cm/\hskip -.25cm\partial z_2)\psi_1^\star  \nonumber \\
f&=&2(f_1^\star f_2-\partial_\mu z_1^\star\partial^\mu z_2)-\overline\psi_1%
\hskip .25cm/\hskip -.25cm\partial\psi_2 -(\overline\psi_2\hskip .25cm/%
\hskip -.25cm\partial\psi_1)^\star.
\end{eqnarray}

\nin We note that 
the transformation of $A_\mu$ has actually the expected form up to
a gauge transformation, which implies that it must be a gauge field. The
fields defined in Eqs.(\ref{defcomp}) are complex: they are not the physical
degrees of freedom but have the generic composite structure that we want to
study. The physical composite degrees of freedom 
are given in \cite{compos}.  There are three $N=1$ scalar
supermultiplets with bosonic part:

\bea
\phi_1&=&\ol\psi_1\psi_2+\ol\psi_2\psi_1\nonu
\phi_2&=&i(\ol\psi_1\psi_2-\ol\psi_2\psi_1)\nonu
\phi_3&=&\ol\psi_1\psi_1-\ol\psi_2\psi_2 \nonu
\eea
which form an SU(2) triplet, and are parity conserving. 
The parity violating $\phi^4 = \ol\psi_1\psi_1+ \ol\psi_2\psi_2$
is discarded here, given the fact that we shall be 
interested in phases where $<\phi_i> \ne 0$, and 
as a result of the Vafa-Witten theorem~\cite{vafa} such spontaneous 
parity violating condensates
are excluded in vector-like theories, such as the ones we are 
interested here.

On the other hand, there are three 
$N=1$ vector supermultiplets (forming again an SU(2) triplet) 
with bosonic part:

\bea\label{vectors}
A_\mu^1&=&2{\cal R}e\left(\ol\psi_1\gamma_\mu\psi_2-z_1^\star\partial_\mu z_2
+z_2\partial_\mu z_1^\star\right)\nonu
A_\mu^2&=&2{\cal I}m\left(\ol\psi_1\gamma_\mu\psi_2-z_1^\star\partial_\mu z_2
+z_2\partial_\mu z_1^\star\right)\nonu
A_\mu^3&=&\ol\psi_1\gamma_\mu\psi_1-\ol\psi_2\gamma_\mu\psi_2\nonu
&&-z_1^\star\partial_\mu z_1+z_1\partial_\mu z_1^\star
+z_2^\star\partial_\mu z_2-z_2\partial_\mu z_2^\star,
\eea
and one SU(2) singlet, which is parity violating: 
\bea 
A_\mu^4&=&\ol\psi_1\gamma_\mu\psi_1 + \ol\psi_2\gamma_\mu\psi_2\nonu
&&-z_1^\star\partial_\mu z_1+z_1\partial_\mu z_1^\star
-z_2^\star\partial_\mu z_2+z_2\partial_\mu z_2^\star.
\label{pviol}
\eea
which, upon taking into account the $\sigma$-model constraint
at a constituent level~\cite{mav} $\sum_{i=1}^{2}z_i^\star z_i 
={\rm const}$, can be rewritten in terms of spinon $J^z$ and holon
currents $J^\psi$ as: 
\begin{equation}
A_\mu^4 = \sum_{i=1}^{2} \left({\ol \psi}_i \gamma_\mu \psi_i + 
2 z_i\partial_\mu z_i^\star \right) \equiv J^\psi_\mu + J_\mu^z 
\end{equation} 
Since in the relativistic theory of the nodal liquids 
we are interested here, the parity violating 
vector gauge field is not allowed to acquire a v.e.v. 
$<A_\mu> =0$, so as not to break the Lorentz invariance of 
the nodal theory, 
the Vafa-Witten theorem~\cite{vafa} is satisfied in this case.
In what follows   
we shall therefore retain  
the parity violating composite excitation, unlike the case of the scalar
$\phi^4$. In fact we shall make supersymmetric spectra by considering 
only $A_\mu^4$, something which we shall justify in the next section
when we discuss in detail the electric charge assignments of the 
various excitations. 
It should be noted that the condition 
$<A_\mu^4>=0$ implies that, on the ground state of our system,
the spinon $J_\mu^z$ and holon $J_\mu^\psi$ 
currents are linked in such a way that only one of them
plays a physical macroscopic r\^ole. This is a welcome feature of the 
spin-charge separation framework.

From the explicit form of the $A_\mu^4$ vector potential 
(\ref{pviol}) we can make the 
interesting observation that this composite excitation has a form similar to 
that one would obtain from a constituent $CP^1$ $\sigma$-model 
coupled to fermions, after elimination of the gauge degree
of freedom:
\begin{equation} 
{\cal S_{\rm CP}}= \frac{1}{\gamma_0} \int d^3x \sum_{i=1}^{2}\left( |(\partial _\mu -iA_\mu^4 )z_i|^2
+ {\ol\psi}_i\left(\br\partial-i\br A_\mu^4\right) \psi_i \right) + \dots 
\end{equation}
As argued in \cite{altaba}, integrating out in this naive continuum model 
the gauge field produces composites of $z$ and $\psi_i$ fermions 
resulting in contact interactions among the fermions. 
This 
indicates that what we have identified as a gauge composite field 
in our case is indeed related to strongly coupled U(1) gauge group 
fields to be integrated out in the much more complicated, but realistic,
Lattice gauge theory~\cite{farakos} 
of fundamental constituents. In such a theory, whose  precise form is unknown,
there are many more non-minimal, higher derivative contact interactions
among spinons and holons, and it is the postulate of supersymmetry
that allows the construction of an effective continuum field theory.   
In realistic situations one may hope that the supersymmetric 
theory of composites would lie in the same universality class
in the infrared as the 
more realistic effective continuum theory representing the true 
low-energy dynamics of the system.

From the three scalar composite supermultiplets, we form one real $N=1$ 
scalar supermultiplet $(\rho,\xi,\sigma)$ 
and one complex $N=1$ scalar supermultiplet $(\phi,\psi,f)$ where: 

\begin{eqnarray}\label{deffields} 
\rho&=&\phi^3 = {\ol\psi}_1\psi_1-\ol\psi_2\psi_2\nonu
\xi&=&(f_1^\star - \hskip .25cm/\hskip -.25cm\partial z_1^\star)\psi_1 + 
(f_1-\hskip .25cm/\hskip -.25cm\partial z_1)\psi_1^\star - (1 \rightarrow 2)\nonu
\phi&=&\phi_1 - i \phi_2 =2{\ol \psi_1}\psi_2 \nonu
\psi&=&\psi'=(f_1^\star-\hskip .25cm/\hskip -.25cm\partial z_1^\star)\psi_2+
(f_2-\hskip .25cm/\hskip -.25cm\partial z_2)\psi_1^\star 
\label{comstr}
\end{eqnarray} 

The $N=1$ composite supersymmetric transformations,
found in \cite{compos}, are then

\begin{eqnarray}\label{transfocompN1}
\delta\rho&=&\overline\varepsilon\xi,~~~~\delta\xi=(\hskip .25cm/%
\hskip -.25cm\partial\rho+\sigma)\varepsilon, ~~~~\delta
\sigma=\overline\varepsilon\hskip .25cm/\hskip -.25cm\partial\xi,\nonu
\delta\phi&=&\overline\varepsilon\psi,~~~~\delta\psi=(\hskip .25cm/%
\hskip -.25cm\partial\phi+f)\varepsilon, ~~~~\delta
f=\overline\varepsilon\hskip .25cm/\hskip -.25cm\partial\psi,  \nonu
\delta A_\mu&=&\overline\varepsilon\gamma_\mu\chi,~~~~\delta\chi = -%
\frac{1}{2}F_{\mu\nu}\gamma^\mu\gamma^\nu\varepsilon =(\partial^\nu
A_\nu-\hskip .25cm/\hskip -.25cm\partial\hskip .25cm/\hskip -.25cm
A)\varepsilon,
\end{eqnarray}

\noindent where $F_{\mu\nu}$ is the \textit{Abelian} field strength
of $A_\mu = A_\mu^4$ (quadratic in the constituent fields), 
and the gaugino $\chi$ is real, given by:

\begin{equation}\label{chidef}
\chi = i(f_1^\star+\hskip .25cm/\hskip -.25cm\partial z_1^\star)\psi_1-i(f_1+%
\hskip .25cm/\hskip -.25cm\partial z_1)\psi_1^\star + (1 \rightarrow 2) 
\end{equation} 

We stress again that above we selected 
the parity violating vector supermultiplet due to 
electric charge assignments to be discussed in the next section. 

\bigskip

In order to elevate the $N=1$ supersymmetry  
to a $N=2$ (Abelian) supersymmetry, we should
couple the complex composite scalar supermultiplet to one of the composite
vector supermultiplets, which we denote as $(A_\mu,\chi)$. 
In \cite{ams} 
we constructed explicitly the covariant 
derivative of $\phi$ by adding to the
complex quadratic scalar field a \textit{higher order (quartic) 
composite} which will
generate the minimal coupling. We took into account the quartic contribution
only and neglected the higher orders: we managed to 
construct a quartic composite
scalar $M$ whose supersymmetric transformation generates the quartic fermion 
$\Lambda$:

\begin{eqnarray}  \label{quartictransfo}
\delta M&=&\overline\varepsilon\Lambda  \nonumber \\
\delta\Lambda&=&\left(-i\hskip .25cm/\hskip -.25cm A\phi+\hskip .25cm/%
\hskip -.25cm\partial M+\mathcal{F}\right)\varepsilon,
\end{eqnarray}

\noindent where $\mathcal{F}$ is a quartic auxiliary field. 
An important point of our construction is that 
the transformation (\ref{quartictransfo}) of the fermion was satisfied
in a gauge defined by the complex equation

\begin{equation}  \label{gaugecond}
\partial^\nu\left(A_\nu\Box\phi\right)=0,
\end{equation}

\nin which implies actually two gauge conditions. This is possible
in (2+1) dimensions, 
since a gauge field has one physical degree of freedom \cite{binegar}. 
Making the substitutions:

\begin{eqnarray}  \label{substitut}
\phi&\to&\Phi=\phi+gM  \nonumber \\
\psi&\to&\Psi=\psi+g\Lambda  \nonumber \\
f&\to&F=f+g\mathcal{F},
\end{eqnarray}

\noindent where $g$ is a dimensionful constant, we  
obtained the expected covariant derivatives

\bea\label{covderiv}
D_\mu\Phi&=&(\partial_\mu-igA_\mu)\Phi\nonu
&=&\partial_\mu\phi-igA_\mu\phi
+g\partial_\mu M +\mbox{sixth order composite}
\eea

\noindent in the (supersymmetry) 
transformation laws of the complex fermion $\Psi$. 

In the above formulae, 
the quartic scalar $M$ that we constructed in order to generate the minimal
coupling $\br A\phi$ is given by

\begin{equation}\label{Mscalar} 
-i\partial^\rho\Box M=\epsilon^{\mu\nu\rho}f\partial_\mu A_\nu
+\partial^\rho\left(2\partial^\mu\phi A_\mu+\phi\partial^\mu
A_\mu+\overline\psi^\star\chi\right).
\end{equation}

\noindent Note that the occurrence of this scalar is specific to 2+1
dimensions, as it is proportional to the topologically conserved
current $J^\rho=\epsilon^{\mu\nu\rho}\partial_\mu A_\nu$, which plays a
central role in the elevation of an $N=1$ supersymmetry to an extended $N=2$
supersymmetry \cite{hlousek}.

On the other hand, the fermion $\Lambda$ is found to have the form \cite{ams}: 
\begin{equation}\label{Lambda1} 
\Lambda = i\left(\Lambda^{(1)} + 2\Lambda^{(2)} + \Lambda^{(3)} +\Lambda^{(4)}\right)
\end{equation} 
where:
\begin{eqnarray} 
\Box\Lambda^{(1)} &=&\partial^\mu A_\mu\psi+\phi\hskip .25cm/\hskip %
-.25cm\partial\chi~\nonu
\Box\Lambda^{(2)}&=&\hskip .25cm/\hskip -.25cm\partial\phi\chi+A_\mu
\partial^\mu\psi~,  \nonumber \\
\Box\Lambda^{(3)}&=&(f-\hskip .25cm/\hskip -.25cm\partial\phi)\chi + (%
\hskip .25cm/\hskip -.25cm\partial\hskip .25cm/\hskip -.25cm A-\partial_\mu
A^\mu)\psi~, \nonumber \\
\hskip .25cm/\hskip -.25cm\partial\Box\Lambda^{(4)}&=&(\partial^\nu A_\nu- %
\hskip .25cm/\hskip -.25cm\partial\hskip .25cm/\hskip -.25cm A)\hskip .25cm/ %
\hskip -.25cm\partial \psi-2f\hskip .25cm/\hskip -.25cm\partial\chi.
\label{Lambda2} 
\end{eqnarray} 

We also constructed 
the complex gaugino $\lambda=\xi-i\chi$, where $\xi$ and $\chi$ (superpartner of $A^4_\mu$)
are respectively defined in Eq.(\ref{deffields}) and Eq.(\ref{chidef}):

\begin{equation} 
\lambda = 2\left(\hskip .25cm/\hskip -.25cm\partial z_2^\star \psi_2
-\hskip .25cm/\hskip -.25cm\partial z_1\psi_1^\star \right)
+ 2\left(f_1^\star \psi_1 - f_2 \psi_2^\star \right)
\label{lambdagaugino} 
\end{equation} 
\nin and finally obtained the
following transformations

\begin{eqnarray}\label{transfofinal}
\delta\Phi&=&\overline\varepsilon\Psi,~~~~~~~~\delta\Psi=(\hskip .25cm/%
\hskip -.25cm D\Phi+F)\varepsilon,  \nonumber \\
\delta\rho&=&\overline\varepsilon\xi,~~~~~~~~\delta
A_\mu=\overline\varepsilon\gamma_\mu\chi,  \nonumber \\
\delta\lambda&=&\left(\hskip .25cm/\hskip -.25cm\partial\rho+\sigma+\frac{i}{2%
}F_{\mu\nu}\gamma^\mu\gamma^\nu\right) \varepsilon  \nonumber \\
\delta \sigma&=&\overline\varepsilon\hskip .25cm/\hskip -.25cm\partial\xi
\end{eqnarray}

\noindent which, when $\varepsilon$ is a complex parameter, constitute a set
of $N=2$ supersymmetric transformations 
for an Abelian Higgs model~\cite{edelstein}.

It should be remarked that the N=2 supersymmetric transformations
are closed in our construction up to terms which, in a lagrangian
formalism, would correspond to higher derivative operators
and hence to irrelevant operators in the infrared (low energies)
in a renormalization-group sense~\cite{ams}. 

Finally, we note that no new constraints arise if we consider higher order
composites: the quartic composite can be seen as the truncation of a series
of composite operators which lead to the exact covariant derivatives if they
are ressummed. The gauge condition (\ref{gaugecond}) gives rise to a gauge
fixing term in the Lagrangian which is also the truncation of a series of
gauge fixing terms.

This completes our review of the construction of an 
Abelian Higgs composite model out of spinon and holon nodal fields. 
At this point we note that in \cite{edelstein}, where the authors consider
`elementary' fields and not composites, the superpartners
transform under an on-shell set of $N=2$ transformations were the equations
of motion of the auxiliary fields are used, in the supersymmetric Abelian
Higgs model. In \cite{ams} we
restricted ourselves only on the algebraic aspects of the composite
supersymmetric transformations and did not 
consider the detailed dynamics of this model, and 
the associated physical consequences in connection with strongly-correlated
electron systems. These topics will be the focus of the following sections.

Notice that, as  a result of our specific construction, by supersymmetrizing 
only one of the vector multiplets, which as we shall see later on is also 
dictated
by electric-charge assignments, we encounter an Abelian Gauge Field 
model.  In addition to the electric-charge argument, 
we also mention that, with the present composite construction,
we were unable 
to generalize the composite superalgebra to an $SU(2)$ supermultiplet 
(constructed out of the parity conserving composites)
since 
too many constraints on the fields would be present: constraints for the
generation of the covariant derivatives and  also for the generation of
the non-Abelian field strengths. It is not clear yet  
whether the parity-violating Abelian structure, generated by  
$A_\mu = A_\mu^4$ could be 
embedded in a non-Abelian group. In fact, as we shall argue below, 
interesting physics already occurs in the case where our Abelian
group 
is {\it non compact}. Some comments on the non-Abelian extension
will be presented as outlook.

\section{Electric charge assignment of Composite excitations} 

An important ingredient, which was so far has been left out from the 
above discussion, is the electric charge (not to be confused
with the statistical charge quantum number discussed so far) assignment of the 
composite excitations discussed above. In view of the spin-charge separation
that characterizes the ground state of the nodal liquid, this issue 
turns out to be quite crucial for the associated physics, and in particular
the superconducting properties of the model. 

At the constituent level, the electric charge assignment of the 
various excitations is well defined~\cite{mav}: spinons, $z_{1,2}$ 
are {\it electrically neutral}, while holons carry electric charge,
which in terms of the (two-component) 
Nambu-Dirac spinors $\psi_{1,2}$ can be given 
the following electric charge assignment: 
\begin{eqnarray}
\psi_1 &=& \left(\begin{array}{c}{\widehat \psi}_1^\dagger 
\\ -{\widehat \psi}_2 \end{array}\right)~ \rightarrow {\rm electric~charge~-e} \nonumber \\
\psi_2 &=& \left(\begin{array}{c}{\widehat \psi}_2^\dagger 
\\ {\widehat \psi}_1 \end{array}\right)~\rightarrow 
{\rm electric~charge~ +e}~, 
\label{elchassign}
\end{eqnarray} 
where ${\widehat \psi}_{i}~, i=1,2$ are Grassmann numbers, denoting
annihilation operators of a hole with non zero electric charge. 
Notice that the above charge assignment are consistent with the 
microscopic structure of the nambu-Dirac spinors in terms of this
Grassmann fields. 
 
In view of this assignment we are now in position to 
discuss the electric charge content of the composite 
operators of the previous section. We commence our discussion
from the 
quadratic parts of these operators. 

First, let us consider the matter parts $\phi, \psi$. From 
(\ref{comstr}) and (\ref{elchassign}), and recalling 
the electrical neutrality of $z_{1,2}$ spinon fields, we do observe that 
$\phi$ has electric charge +2e, while $\psi$ has electric charge +e.

In a similar spirit, the gauge field $A_\mu=A_\mu^4$  (\ref{pviol}),
is electrically neutral, and 
so is $A_\mu^3$ in (\ref{vectors}). 
On the other hand, we remark that the vector fields $A_\mu^{1,2}$ 
would not have 
definite electric charge. This is one of the reasons why we form 
the supersymmetric composite theory in terms of the $A_\mu^4$ field. 
The bosonic partner fields $\rho=\phi^3$ and $\phi^4$ are 
electrically neutral.

The reason why we do not choose the vector $A_\mu^3$ rests with the 
requirement of a definite electric charge for the gaugino of the 
N=2 multiplet $\lambda$ (\ref{lambdagaugino}). 
First of all we observe that, as it stands, the gaugino $\lambda$ 
does not have a definite electric charge, unless the auxiliary 
fields of the constituent scalar multiplets 
\begin{equation} 
f_1=f_2=0~.
\label{auxfieldconst}
\end{equation} 
This is a question that depends on the details of the constituent 
dynamics. For instance, (\ref{auxfieldconst}) 
characterises the Wess-Zumino model, and some supersymmetric 
gauge theories. However, in the case of the supersymmetric constituent model of \cite{mav}, 
the dynamics is that of a $CP^1$ $\sigma$-model, which has the well known 
constraints $\sum_{a=1}^2{\ol Z}_aZ_a=1 $ in superfield language.
Such constraints imply the on-shell condition 
$f_1=f_2 = {\ol \psi}_1 \psi_1 +{\ol \psi}_2 \psi_2 
=\phi ^4$. In our theory we may discard the parity violating condensate 
$\phi^4$, and hence impose the constraint that on physical states 
(\ref{auxfieldconst}) holds, implying that the physical gaugino excitation
$\lambda$ (\ref{lambdagaugino}) would have definite electric charge +e.

\begin{figure}[t]
\epsfxsize=4in
\bigskip
\centerline{\epsffile{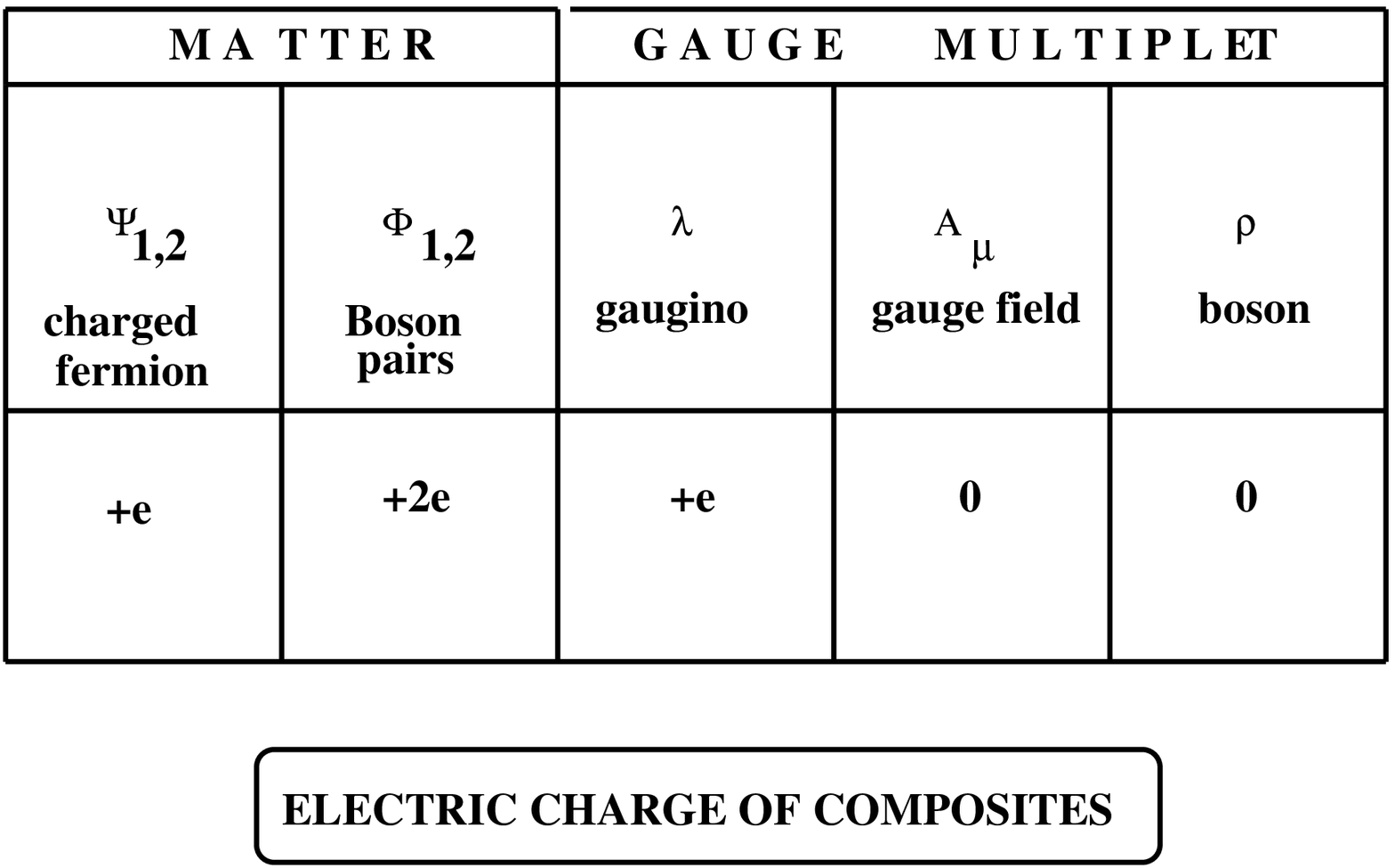}}
\caption{The electric charge assignments of the various composite 
excitations of the supersymmetric Abelian Higgs model.
\label{fig:electable}}
\end{figure}

Notice that the particular construction of the gaugino by combining 
the superpartners of the (parity conserving) excitation $\rho = \phi^3$ and 
the (parity violating) vector field $A_\mu^4=A_\mu$ 
is the only one that leads to a $\lambda$ field that has definite
electric charge. Indeed, if the corresponding gaugino $\lambda$ 
were constructed as in (\ref{lambdagaugino}) but 
from the superpartners of $\rho = \phi^3$ and $A_\mu^3$ 
it would not have a well defined
electric charge. 
Notice that the choice of $A_\mu^3$ in the construction of the 
gaugino would necessitate the involvement of the parity violating 
$\phi^4$ condensate, which could not acquire a Vacuum Expectation 
Value according to the Vafa-Witten theorem~\cite{vafa}. Hence this case
is also excluded, leaving the current scenario as a unique one
with the desired physical properties to be discussed in later sections. 

We next come to the quartic parts of the composite operators. 
With the above electric-charge assignments, we do observe that 
$\Lambda$ (\ref{Lambda1}),(\ref{Lambda2}) 
has a definite electric charge +e provided that the 
quadratic composite auxiliary field $f$ vanishes:
\begin{equation} 
f =0
\label{auxfieldconst2}
\end{equation} 

There is a small but important point regarding the 
the electric charge of the scalar $M$ (\ref{Mscalar}).
In our N=1 construction,  
due to the ${\ol \psi}^\star\chi$ term, 
the scalar $M$ seems not to have a definite charge.
However, as remarked after Eq.(\ref{transfofinal}), the elevation of 
the supersymmetry to 
the extended N=2 case, requires the relevant parameter
$\varepsilon$ to become complex. This allows an analytic continuation of the
real spinor $\chi$ to a complex one, which may be chosen in such a way that 
it has a (global) electric
charge $-e$ like. The correct assignment then of electric charges (associated
with the definition of the proper global rotations of spinors) 
is such that 
in the expression for $M$ (\ref{Mscalar}) $\chi \to \chi^\star$, while
the expression for $\Lambda^{(3)}$ stays as in (\ref{Lambda2}).
It turns out that this assignment is consistent with N=2 supersymmetry,
and also with the conservation of electric charge in the various
terms of the Lagrangian that describes the relevant dynamics, to be 
discussed in the next section. 

With this assignment. 
all the 
quartic parts of the scalar have charge +2e in accordance with the 
quadratic parts, which allows the assignment of charge +2e to 
the scalar $\Phi$ of the matter multiplet. 
Notice the different electric charge assignments between the 
boson and fermions of the matter supermultiplet, which is in 
accordance with the expected explicit breaking of the supersymmetry 
upon coupling to real electromagnetism. 

A summary of the electric charge assignments of the various 
composite excitations of the supersymmetric model 
is given in figure \ref{fig:electable}.

\section{Application to Strongly Correlated Electrons}

\begin{figure}[t]
\epsfxsize=2.5in
\bigskip
\centerline{\epsffile{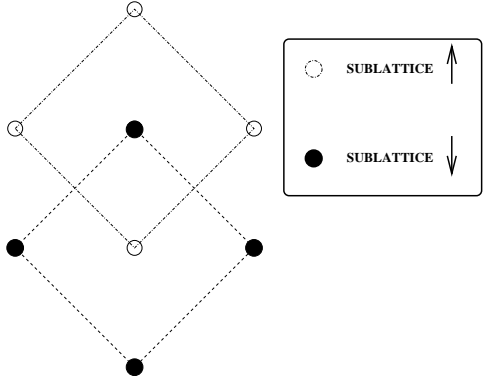}}
\caption{An antiferromagnetic sublattice in the scenario
of \cite{dorey}; sublattice hopping (dashed, or dashed-dotted lines)
defines effectively two 
kinds of holons, with opposite
statistical gauge couplings, which effective hop within a single sublattice, 
without 
direct intersublattice hopping; 
mixing of the sublattices is only allowed via 
gauge field frustration, described in \cite{dorey}. 
The holon excitations can be taken to be linear excitations
of a single nodal point in the corresponding $d$-wave fermi
surface of the high temperature superconductor. 
However, this construction  
is  {\it not} 
the appropriate model to be considered in the non-Abelian 
spin-charge separation ansatz (\ref{ansatz}) 
of \cite{farakos,mav}, adopted here, 
where direct intersublattice hopping is allowed.  
It is only mentioned here
for instructive purposes in order to understand which way the continuum limit
should be taken in our case. 
\label{fig:sublattice}}
\end{figure}

\paragraph{}
\underline{The Effective Continuum Lagrangian}

\paragraph{} 

\begin{figure}[t]
\epsfxsize=4in
\bigskip
\centerline{\epsffile{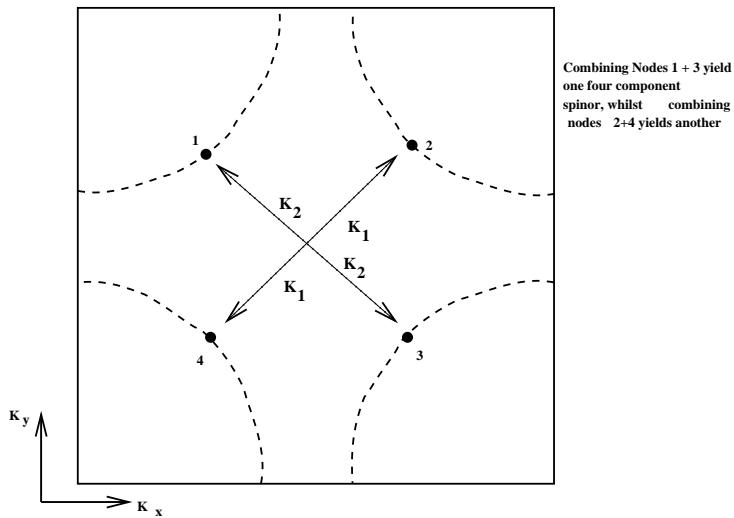}}
\caption{The fermi surface of underdoped cuprates consists of just four nodes.
The dashed line is the putative fermi surface.
The continuum effective theory 
may be obtained by linearisation about such nodes,
which leads, at the constituent level, to two flavours of 
four-component Dirac spinors for the holon degrees of freedom,
each flavour being obtained 
by combining the nodes along the diagonal 
as indicated in the figure (which is a standard 
procedure in deriving the continuum limit of lattice gauge theories
in particle physics).
\label{fig:nodes}}
\end{figure}

In this section we shall consider the physical application of the 
above consideration in the phase diagram of the nodal liquid. 
First of all we remind the reader that the above theory
assumed the existence of nodes in the Fermi surface of 
the system, and concentrated in linearising around such nodes.
In terms of realistic situation, with possible relevance to high 
temperature superconductors, we are working on a temperature-doping
diagram on regimes where such nodes are present. 
This is the case of low doping (underdoped cuprates), 
where the fermi surface consists of four nodes (as in figure \ref{fig:nodes}),
or in the optimal
doping superconducting case, where the $d$-wave superconducting gap
(which is an experimental fact) has nodes as well.  

Linearising the lattice theory of the 
{\it constituent} excitations (spinons and holons)
about such nodes implies the usual 
lattice doubling, that is the presence in the continuum
relativistic theory of {\it four component} constituent spinors (holons).
There are different ways of obtaining the continuum limit, all of which 
may not be physically equivalent. Let us briefly describe them,
which will be instructive in deciding on the appropriate one 
to be adopted in our case here.

One way of obtaining the continuum limit 
at a constituent level
is outlined in \cite{semenoff,dorey}, 
and is appropriate for considering
excitations about a {\it single node} of the fermi surface.  
Consider a 
planar lattice (lying, say, on the spatial $x-y$ plane),
where the original microscopic
antiferromagnetic model lives on. In the model of \cite{dorey} 
direct intersublattice hopping is forbidden, and the holons are confined
within each sublattice, thereby defining two species (`colors') of holons.
This model is appropriate for the Abelian spin-charge separation
ansatz where an electronic degree of freedom in each sublattice
is split into $c_{i,\alpha} =z_{i,\alpha}\psi_{i}^\dagger$, 
where $\alpha=1,2$,  
and $\psi$ are spinless Grassmann numbers (holons),
while $z_\alpha$ (magnons) are boson doublets~\cite{dorey}. 
It is only $\psi$ that carries a sublattice `color' index. 

Each subattice is divided into two sublattices according to the 
hopping of holons (see figure \ref{fig:sublattice}).   
Then an arbitrary sublattice point, 
corresponding to the 
position vector ${\vec i}$ reads ${\vec i} =n_x {\hat a}_x + n_y {\hat a}_y $,
where $n_{x,y}$ are positive integers (including zero), and 
${\hat a}_{x,y}$ are orthogonal unit vectors of the sublattice. 
We divide spinor components on each sublattice 
into four kinds, according to the {\it parity} of the 
integers $n_x$ and $n_y$ as follows: 
\begin{equation} 
\psi _1 = ({\rm even}, {\rm even})~, ~ 
\psi _2 = ({\rm even}, {\rm odd})~, ~ 
\psi _3 = ({\rm odd}, {\rm even})~, ~ 
\psi _4 = ({\rm odd}, {\rm odd})~,  
\label{doubling1} 
\end{equation} 
where $\psi $ are Grassmann numbers,  
and one constructs four-component spinor (of each sublattice `color') 
about a {\it single node} 
on the fermi surface as follows: 
\begin{equation} 
\chi^{(4)} = \frac{1}{\sqrt{2}}\left(\begin{array}{c} \psi_1 + \psi_4 \\
\psi_2 + \psi_3 \\ \psi_2 - \psi_3 \\ \psi_1 - \psi_4 \end{array}\right)
\label{doublingspinor}
\end{equation}
The linearisation 
about a single node on the fermi surface, then, is done as in 
\cite{dorey} by considering the Fourier components of the above 
spinors; in this way, as one takes the continuum limit, 
by superimposing the two sublattices, 
the Lattice doubling yields four component
Dirac spinors, around each node, and the sublattice structure provides
two colours of such spinors. The Dirac nature is obtained by assuming 
a flux $\pi$ statistical gauge background per 
sublattice plaquette~\cite{dorey}.
In two component notation, then,
one obtains two `colours' and two flavours of four component 
Dirac spinors (electrically charged). 

The above construction is not appropriate, however, for 
our considerations here, which are concerned with the non-Abelian 
spin-charge separation ansatz of \cite{farakos,mav} defined on
a planar antiferromagnetic lattice but allowing intersublattice 
hopping. The ansatz was introduced for  
a particle-hole symmetric formulation away from half-filling and reads:
\begin{equation} 
\chi _{\alpha \beta }\equiv \left(\begin{array}{cc}
{\psi _1} \qquad {\psi _2} \\
 {-\psi _2^\dagger} \qquad {\psi _1^\dagger}\end{array} 
\right)_i\left( \begin{array}{cc}  
 {z_1} \qquad {-\bar z_2} \\ 
 {z_2} \qquad {\bar z_1}\end{array} \right)_i\,
\label{ansatz}
\end{equation} 
where
the fields $z_{\alpha,i}$ obey canonical {\it bosonic}
commutation relations, and are associated with the
{\it spin} degrees of freedom (`spinons'), whilst the fields $\psi$
are Grassmann variables on the lattice, which obey fermi 
statistics, and are associated with the electric charge degrees of freedom
(`holons'). The ansatz admits hidden non-Abelian local $SU(2)$ 
spin symmetries at the constituent level, as discussed in our 
previous works~\cite{farakos,mav,sarbdual}. 
The important point to notice is that now the holon degrees
of freedom do carry a spin index. 
As shown in detail in \cite{mav}, from this anstaz one can construct 
the two Dirac spinors (\ref{elchassign}), which play the r\^ole of 
the effective electrically-charged degrees of freedom for the problem,
with the electric charge assignment of the previous section.

An important question arises at this point 
as to how one should take the continuum limit of this 
non-Abelian spin-charge separation theory. 
We propose to follow the procedure of \cite{herbut}
appropriate for BCS superconductivity, despite the fact that 
our proposed mechanism for superconductivity, as we shall see below,  
will {\it not } be 
the conventional BCS, but that of \cite{dorey}. 
In \cite{herbut} 
the four component spinors of the continuum theory 
are obtained by {\it combining
nodes} along the diagonal as indicated in figure \ref{fig:nodes}. 
This is dictated by the starting point which is a BCS-like pairing 
interaction (at finite temperature $T \ne 0$) 
among BCS-like quasiparticle 
excitations near nodes at opposite ends of the 
diagonals of figure \ref{fig:nodes}~\cite{herbut}:
\begin{equation} 
{\cal S} = T \sum_{{\vec k},~\alpha,~\omega_n}
[(i\omega_n - \xi_{\vec k})c_\alpha^\dagger ({\vec k},\omega_n)
c_\alpha ({\vec k},\omega_n) - \frac{\alpha}{2}\Delta ({\vec k}) 
c_\alpha^\dagger ({\vec k},\omega_n)
c_{-\alpha}^\dagger (-{\vec k},-\omega_n) + h.c. + \dots] 
\label{bcspairing}
\end{equation} 
in a standard notation, with $\alpha=1,2$ denoting the spin up and down 
states of the third component of the 
electron spin, $\omega_n$ being the fermionic Matsubara frequencies, 
$\Delta ({\vec k})$ being the $d$-wave gap, and the $\dots$ indicating
other possible short range interactions between quasiparticles.  
In the approach of \cite{herbut}, where spin-charge separation is
not considered explicitly, $c_\alpha,c_\alpha^\dagger$,
$\alpha=1,2$ are electron operators,
while in our approach here and in \cite{mav}, $c_\alpha,c_\alpha^\dagger$
should be replaced by the corresponding holon operators/Grassmann numbers 
$\psi_\alpha,\psi^\dagger_\alpha$, $\alpha=1,2$ 
appearing in (\ref{elchassign}).
It must be noted that this way of combining the four nodes in order to form 
a continuum theory from a lattice one 
is also 
the standard procedure of taking the continuum limit in 
lattice particle theory models. 

If one adopts the starting point (\ref{bcspairing}) for the constituent
theory, 
then there are two ways of constructing 
the four-component continuum constituent spinors, 
corresponding to holons in our case, 
as discussed in \cite{herbut}~\footnote{However, 
it should be stressed 
that the associated physics in those works  
is entirely  different
from that of our model.
In contrast to our case, in \cite{herbut} 
the authors do not consider direct spin-charge separation. 
Moreover their effective 
spinons are viewed as electrically neutral fermions,
while the holons are electrically charged
bosons. 
Of course this by itself would not lead to physical 
differences, given that 
there is a physical equivalence 
between the two formalisms
in view of bosonisation~\cite{marchetti}. But the important difference
is the fact that these workers use pure QED3 to describe frustration
of holes. In \cite{dorey}, and in all of our works so far on the subject,
and here, the important point is that we use QED coupled with 
opposite statistical charge to the two species. In \cite{dorey}
we have discussed the difference if one used pure QED as in \cite{herbut}. 
In that case the anomalous graph would imply chiral symmetry breaking.
It is the opposite-coupling QED model of \cite{dorey}, which implies
unconventional superconductivity. 
In contrast, in \cite{herbut},  
as far as the proposed mechanisms for superconductivity are concerned, 
the authors assume that the superconducting state, apart 
from being a $d$-wave, exhibit otherwise the standard BCS phenomenology.},
leading possibly 
to different physics as far as the properties of the insulating 
phase are concerned (to which we are not directly concerned
in this article). One (Herbut in \cite{herbut}) 
is to define the four-component 
spinors in each pair of nodes as:
\begin{equation} 
\Psi ({\vec q},\omega_n)_{i=1,2}= \left(c^\dagger_1 ({\vec k}, \omega_n), ~
c_2(-{\vec k}, -\omega_n),~ c^\dagger_1 ({\vec k}-{\vec Q}_i, \omega_n), ~
c_2(-{\vec k}+{\vec Q}_i, -\omega_n)\right)~, 
\label{four1}
\end{equation}   
where ${\vec Q}_i=2{\vec K}_i~$
is the wave vector connecting the nodes 
within the diagonal pair $i=1,2$ (c.f. figure \ref{fig:nodes}), 
and ${\vec q}$ 
is measured from the (putative) fermi surface ${\vec k}={\vec K}_i + {\vec q}$,
$|{\vec q}| \ll |{\vec K}_i|$ for the pair $i=1,2$. 
In our case, of course, the electron
operators should be replaced by the Grassmann number operators
$\psi_{1,2}$, defined in (\ref{elchassign}),  
as mentioned previously.

The alternative way of forming the four-component continuum spinors,
within the context of BCS-like quasiparticle pairing interaction 
(\ref{bcspairing}) is described 
in Balents {\it et al.} in \cite{herbut}, where the four-component
spinors are constructed as: 
\begin{equation} 
\Psi ({\vec q},\omega_n)_{i=1,2} = 
\left(c^\dagger_1 ({\vec k}, \omega_n), ~
c_2(-{\vec k}, -\omega_n),~ c^\dagger_2 ({\vec k}, \omega_n), ~
-c_1(-{\vec k}, -\omega_n)\right)~,
\label{four2}
\end{equation}  
again for each pair of nodes $i=1,2$.
Again, within our spin-charge separation framework, the electron operators
should be understood as representing the Grassmann number fields $\psi_{1,2}$.

The main difference between the two approaches lies in the ability to describe
properly the insulating phase, which is of no concern to us here. 
Formally, the $\gamma$-matrix and the emergent gauge symmetry structure 
of  
the second approach is that of {\it spin rotations}. It is 
this approach which we 
adopted in our formalism in \cite{mav},
and will be adopted here as well. 
As can be readily seen from 
(\ref{elchassign}),(\ref{four2}), 
in our case the construction of four-component spinors is obtained  
by combining appropriately 
the two component Dirac spinors (\ref{elchassign}) at the nodes
along the diagonals of figure \ref{fig:nodes}. 
As mentioned 
already, the crucial physical 
difference of our analysis from that in \cite{herbut}
is that here the four component spinors are electrically charged
(as in (\ref{elchassign})) and represent holon and not electron
degrees of freedom.
The Dirac nature of the corresponding constituent spinor lagrangian, 
is again obtained 
upon coupling 
the consituent 
holons to a background statistical gauge field 
with flux $\pi$ per lattice plaquette.

Before proceeding further an important comment is in order. In \cite{mav} 
one could have taken the four component continuum limit spinors
around a given node by simply combining the two two-component
spinors in (\ref{elchassign}) into a {\it single four 
component one}, with simply taking the Fourier components
of the corresponding Grassmann quantities. 
In this case one can obtain a perfectly well-defined continuum 
supersymmetric gauge theory at a constituent level in certain regime 
of parameters of the microscopic model, as discussed in 
\cite{mav}, around {\it each node}. The construction will then lead to 
four replicas of such theory for the four nodes. 
At a composite level, however, such a construction would not 
yield the second species of chiral matter multiplets 
(c.f. $Q_{1,2}$ in a superfield notation below and in Appendix), 
which one needs in order to construct a {\it parity-invariant}  
effective composite theory. 
This opportunity is provided to us by the above-mentioned  
construction
of the continuum effective theory, adopted here, which combines
the pair of nodes along the diagonal of figure \ref{fig:nodes}.

At the constituent level, therefore, 
we form two four component continuum spinors by combining 
lattice fermionic composite excitations near  
nodes lying along the diagonal of figure \ref{fig:nodes}. 
The important point to notice is that each diagonal pair 
of nodes yields two two-component complex Dirac spinors in the continuum
and thus {\it four} constituent fermionic degrees of freedom.
For our purposes here,
the composite supermultiplets will then be constructed in the 
{\it continuum limit} out of the continuum constituent degrees 
of freedom available to us.  
Each pair of nodes then implies one composite supermultiplet
at supersymmetric points, since the latter is made out of 
four constituent fermionic 
degrees of freedom. We thus finally obtain two composite 
matter supermultiplets $Q_{1,2}$ in supefield notation 
(c.f. Appendix and below).
It is important to notice that in our construction 
we consider the composite gauge field excitation, $A_\mu^4$, and its partner 
under supersymmetry (gaugino), 
as expressing interactions
(frustration) among those pairs of nodes. 
Note that the complex gaugino of the N=2  
gauge supermultiplet keeps track of the correct doubling of degrees 
of freedom in the continuum limit, thereby providing another
physical reason for the physical relevance of the N=2 supersymmetry, 
in addition to the electric charge assignments discussed previously.

A question arises at this point as to the relative sign of the statistical
gauge coupling among the two pairs of the diagonal nodes.
As a result of energetic arguments in favour of a parity 
conserving ground state~\cite{vafa}~\footnote{Caution is expressed
at this point, though, regarding the lattice composite theory where 
the Vafa-Witten theorem 
may actually fail~\cite{farakos}
due to the existence of lattice operators that violate parity
(proportional 
to the lattice spacing, and thus corresponding in the 
continuum limit to operators with 
higher derivatives than the ones considered here). 
The issue is whether such lattice parity-violating operators
can be relevant in a renormalization group sense. From a naive point 
of view, since such operators correspond to higher derivative operators
in the continuum,  
they are expected not to affect the infrared universality class
of the model. 
But this expectation may be naive, and a precise 
renormalization group analysis 
needs to be performed. Such issues will not concern us here.}, 
we observe that at 
the supersymmetric 
points parity conservation actually
necessitates 
{\it opposite} couplings with the statistical gauge field
between the two pairs of nodes of figure \ref{fig:nodes}.
This is due to the Yukawa type coupling of the gaugino with the 
members (fermion and boson) of the matter multiplet, 
$\Phi^\star \ol \lambda \Psi$, which 
would be otherwise parity violating as becomes clear from the 
relevant Lagrangian terms shown below.  Notice that such Yukawa type terms 
are not present in non supersymmetric theories, and hence the
arguments on opposite couplings among the diagonal nodes 
is an exclusive feature of the existence of supersymmetric points. 

The resulting dynamics at the supersymmetric
points of the parameter space of the spinon-holon 
composite system 
may then be summarized by that of the N=2 supersymmetric 
Abelian-Higgs model or, N=2 Supersymmetric Quantum
Electrodynamics (SQED), whose properties, including 
the various exact results on its phases, are reviewed in the 
Appendix. For our purposes in this section we give the on-shell
supersymmetric Lagrangian describing the dynamics
of the nodal liquid at supersymmetric points: 

\bea\label{onshelllagrang2}
{\cal L}_{on-shell}&=&-\frac{1}{4}F_{\mu\nu}F^{\mu\nu}+\frac{1}{2}\partial_\mu\rho\partial^\mu\rho+
\frac{i}{2}\ol\lambda\br\partial\lambda\\
&&+\hf D_\mu\Phi_1\left(D^\mu\Phi_1\right)^\star+
\hf D_\mu\Phi_2\left(D^\mu\Phi_2\right)^\star
+\frac{i}{2}\ol\Psi_1\br D\Psi_1+\frac{i}{2}\ol\Psi_2\br D\Psi_2\nonu 
&&+g\frac{i}{2}({\Phi}_1^\star\ol\lambda^\star\Psi_1-{\Phi}_2^\star\ol\lambda^\star\Psi_2 - \mbox{c.c.})
-\frac{g}{2}\rho(\ol\Psi_1\Psi_1-\ol\Psi_2\Psi_2)-U(\rho,\Phi_1,\Phi_2),\nonumber
\eea

\nin where $(\Phi_1,\Psi_1)$, $(\Phi_2,\Psi_2)$ 
are the two sets of superpartners (corresponding to chiral 
matter superfields  $Q_{1,2}$
c.f. Appendix),  
associated with the two pairs of nodes of figure \ref{fig:nodes}, 
and 

\bea
D_\mu(...)_1&=&\partial_\mu(...)_1+igA_\mu(...)_1,\\
D_\mu(...)_2&=&\partial_\mu(...)_2-igA_\mu(...)_2,\nonu
U(\rho,\Phi_1,\Phi_2)&=&\frac{g^2}{2}\rho^2(|\Phi_1|^2+|\Phi_2|^2)+
\frac{g^2}{8}(|\Phi_1|^2-|\Phi_2|^2)^2.\nonumber
\eea

From a physical point of view
it must be stressed that the gauge field here is not 
the real electromagnetic field, but the statistical one,
associated with spin and hole frustrations. 
We shall consider the coupling of 
the external electromagnetic field later on.

\begin{figure}[t]
\epsfxsize=3in
\bigskip
\centerline{\epsffile{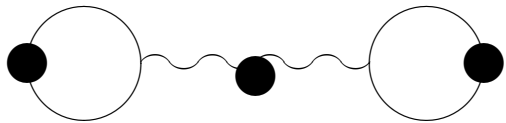}}
\caption{The current-current diagram which leads to massless pole and
superconductivity, via the anomaly mechanism of ref. \cite{dorey}.
The wavy lines represent the statistical photon (from the N=2 vector
multiplet of SQED), and the 
continuous lines are fermions (from the chiral matter
multiplet). Each blob at the two fermion loops
indicate an insertion of the fermion current operator with respect to the 
fermions in the chiral matter multiplet of SQED.
\label{fig:sc}}
\end{figure}

The diagram of figure \ref{fig:sc} is the Landau criterion for 
superconductivity, the electric current-current correlator proceeding via the 
anomaly mechanism of \cite{dorey}. 
In view of the SQED lagrangian (\ref{onshelllagrang}), 
which describes our on shell continuum effective dynamics of the nodal liquid
at supersymmetric points, 
the statistical photon $A_\mu$ can couple to the 
electrically-charged fermions of the chiral multiplet.

We should stress at this point 
that all the considerations in this section
will pertain to the {\it continuum composite theory}. 
On the lattice the composite theory may differ from
its continuum counterpart by various field operators. 
Our {\it conjecture} is that
the two theories belong to the same {\it renormalization group universality class},
in other words the operators by which they may differ are irrelevant
in a renormalization-group sense. This remains to be shown rigorously
by performing a detailed renormalization group approach 
to the continuum limit, which however lies beyond the scope of 
the present article. 

\paragraph{} 

\underline{Coulomb Phase implies Superconductivity} 

\paragraph{} 
The Coulomb phase of the N=2 SQED 
is 
characterized by a non zero $<\rho>$, acting as a mass term for the 
fermions $\psi$ (fermion mass gap). 
From the point of view of the constituent theory
this is the phase where the original holons have a parity 
invariant condensate $\phi^3={\ol \psi_1} \psi_1 - {\ol \psi_2 }\psi_2$
which, for instance, can be generated by the strongly coupled U(1)
of the constituent theory~\cite{farakos} that lead to the composite picture.
In this massive phase, 
the fermions lead to a non zero value of the anomalous one-loop graphs 
of fig. \ref{fig:sc}, in the way explained in \cite{dorey}. 
If the photon is massless this leads to a 
massless pole in the zero-temperature correlator and hence 
to superconductivity. 
The presence of monopole plasma phases destroys this pole,
but this monopole phase is absent in the non compact
Abelian Higgs model discussed here. 
Hence we can safely identify the Coulomb phase of the N=2 SQED
with the superconducting phase of the nodal liquid. 

The local order parameter $<\rho> \ne 0$ is electrically neutral,
but as discussed in \cite{dorey} there is electrical pairing,
which manifests itself upon coupling the theory to an external
(true) electromagnetic field. The pairing manifests itself through 
the so-called mixed Chern-Simons term of the effective theory obtained
after integrating out the (massive) fermionic degrees of freedom
in the parity-invariant theory:

\begin{equation} 
{\cal S}^{eff} \ni \frac{ie}{\pi}\int d^3x \epsilon^{\mu\nu\rho}A_\rho
\partial_\nu {\cal A}^{\rm ext}_\mu 
\label{mixed}
\end{equation} 

\nin where $A_\mu$ is the statistical gauge field, and ${\cal A}_\mu^{\rm ext}$
is the externally applied  electromagnetic potential, due to the electric 
charge of the holons. 

Consider the case of a compact statistical gauge field, which undergoes
large gauge transformations. The situation at any finite 
(no matter how small) temperature $\beta^{-1}$
(so that the time direction is compact) is then the following: 
the large gauge transformation $A_\mu \to A_\mu + \partial_\mu \Omega$,
with $\Omega(\beta) = \Omega (0) + 2n\pi$, $n$ being the winding number of the 
topological class in which the gauge transformation is, 
and $\Omega$ constant on any two dimensional space, 
one obtains for a static external field \cite{dorey} the following change of the mixed 
Chern-Simons term (\ref{mixed}):

\begin{equation} 
\delta {\cal S}^{\rm eff} =2ine \int_{\rm space} d^2x 
\epsilon_{ij}\partial_i {\cal A}_j^{\rm ext}=2ine{\cal F},
\label{change}
\end{equation}

\nin where ${\cal F}$ is the magnetic flux. 
Consider now a superconducting annulus (Corbino disc geometry), as standard
for a pairing demonstration in superconductors. 
Due to the Meissner effect, which has been rigorously demonstrated
for this case in \cite{dorey}, the total magnetic flux ${\cal F}$ through a
region bounded by a closed path winding once ($n=1$) round the origin in the interior
of the sample, must then be quantized as \cite{dorey} (Plank's constant $h$ and the speed 
of light $c$ are restored here)

\begin{equation} 
{\cal F} = n'\frac{hc}{2e}~, \qquad n'={\rm integer}
\label{fluxq}
\end{equation} 

\nin indicating pairing +2e, as observed experimentally. Clearly 
the respective pairs could be formed out of 
condensates of fermions $\Psi$, or holons alone, 
which have such a charge +2e. In terms of four component spinors
a pairing condensate could have been $<\Psi \gamma_5 \Psi>$, where $\gamma_5$ 
is a $4\times 4$
chirality matrix that belongs to a 
reducible representation of the Dirac algebra in (2+1)-dimensional theory,
being appropriate only for theories with even number of continuous 
two-component spinors. This is always the case of lattice models due to 
doubling. 

There is an issue as to the validity of having such condensates
still compatible with supersymmetry.
We think that 
a way out of this issue 
is  the fact that the pairing mechanism described above
pertains strictly only in the presence of an external electromagnetic field,
which breaks supersymmetry explicitly. 
It should also be born in mind that 
such condensates would break 
the statistical gauge invariance as well (spontaneously).  

We also stress that such a pairing in the superconducting 
phase occurs in our model 
only in the {\rm compact} SQED case, in which, as we shall discuss below, 
and reviewed in the Appendix, 
there are monopole (singular) configurations of the statistical gauge field
$A_\mu$. This is a unique feature of our model which differentiates
it from other approaches in the literature, 
such as that of \cite{herbut} {\it etc.}
The important point with the superconductivity scenario of \cite{dorey}
is that the relevant order parameter $<\rho>$ {\it does not break any
symmetries} of the original lagrangian, and hence, at least to 
all orders in perturbation theory, the superconductivity should be 
considered of Kosterlitz-Thouless type, where strong phase fluctuations
suppress the coherence of its phase. This is an important fact of the
nodal liquid superconducting phase, and this lack of phase coherence
is crucial in interpreting (as we shall see later on) the 
monopole plasma phase of the Coulomb branch in the case of compact SQED 
as corresponding to a pseudogap phase. 

The above scenaria for superconductivity occur strictly speaking for 
zero temperature. At any finite temperature, there is a plasmon thermal
mass for the longitudinal component of the gauge field, and moreover
the supersymmetry is explicitly broken. 
However, upon coupling the system to an external magnetic field
there is still screening of the magnetic field lines 
up to a given critical temperature, and hence one can still speak about
superconducting properties, as commented upon in \cite{dorey}.

\paragraph{} 

\underline{Higgs Phase and Pseudogap} 

\paragraph{} 
In the Higgs phase the gauge field is massive, and hence there is no 
pole in the graph of fig. \ref{fig:sc}, and hence no superconductivity
or phase coherence. This is a phase, however, which is characterised by the 
presence of parity conserving condensates of the field $\phi$ which have 
electric charge 2e. Since $<\rho>=0$ in this phase the condensates
are massless (gapless), 
however their presence implies pairing of holons (charge +2e).
Again, in view of the fact that (in the absence of an external 
electromagnetic field) the condensates do not break any of the 
original symmetries of the Lagrangian, probably implies a 
Kosterlitz-Thouless like non superconducting pairing. 

We therefore propose to identify this phase with a pseudogap phase for the 
nodal liquid. Since in this phase there are no monopoles of the 
statistical gauge field there will be no stripes here, so this phase
would be a non-striped regime of the pseudogap phase. 

\paragraph{} 
\underline{Compact SQED and Stripe Phase}

\paragraph{} 
We now turn to a detailed study of the case where the gauge group is 
compact, as seem to be necessitated by electric pairing arguments
in the superconducting case, mentioned above.
In the compact case, as discussed in \cite{strassler}, and reviewed
briefly in the appendix, one has monopole (singular) configurations of the 
gauge field $A_\mu =A_\mu^4$. From (\ref{pviol}) we observe that 
this can indeed be the case if the constituent theory of spinon and holons,
the $CP^1$ supersymmetric $\sigma$-model, lives in one of its non-trivial
topological sectors. Singularities in the spinon and/or holon current
may indeed exist, and such singularities would necessarily imply a compact
gauge field. Compact gauge fields may also be expected in general
for lattice gauge theories, which is the case of the microscopic theory
under consideration here. As we have discussed previously such compact
statistical gauge fields are necessary for electric flux quantization,
characteristic of electric pairing in the superconducting phase.

{}From (\ref{pviol}) and the connection of the $A_\mu^4$ 
field with that of a $CP^1$ $\sigma$-model like field 
at a constituent level, as discussed previously, 
we observe that the monopole configurations (singularities)
could come from the $z$ (spinon) sector, 
in the way discussed in detail in 
\cite{polonyi}. The only extension in our case is the fact that
there are monopole configurations in each species, i.e. in each 
pair of nodes obtained by the combination of figure \ref{fig:nodes}. 
These monopole configuration are identical between the nodes 
by symmetry upon the exchange of nodes~\footnote{In fact it is
this symmetry argument that cancels out any trace of such monopole
singularities in the parity-conserving vector composite $A_\mu^3$ 
(\ref{vectors}), thereby singling out the $A_\mu^4$ as the appropriate
compact gauge field of this framework.}.

The monopoles could form a stable monopole phase, which is true in 
supersymmetric theories as well, with the exception of the SU(2) 
N=2 model of \cite{affleck}, to a discussion of which we shall come 
later. 
The presence of stable monopole phases contributes
non-trivially to the generation of a superpotential 
for the theory $W$, whose form depends on the fugacities of the 
monopoles in the way explained in the Appendix. The monopoles
form a plasma phase~\cite{polyakov}, and 
are responsible for resulting in a massive dual of the statistical 
photon, which will destroy the superconductivity due to 
the anomalous graphs of figure \ref{fig:sc}. This feature is 
also valid in the SQED case, except in the case where SQED is just
embedded in a supersymmetric $N=2$ SU(2) case, in which case there 
is no stable monopole phase~\cite{intril,deboer,strassler}. 
This is discussed briefly in the Appendix. 

Another way of seeing the incompatibility with the 
stable monopole phase may be inferred 
from the absence of any flux quantization
as follows~\cite{mavromatos}: when consider the quantum fluctuations
of isolated monopole configurations, as is the case of a stable 
monopole plasma phase, it is necessary to abandon in a 
path integral fixed boundary conditions for the large 
gauge transformations and instead consider free ones. This is 
essential for the mathematical consistency of the theory,
as argued in \cite{palla}. Free boundary conditions imply that 
in a path integral one has to integrate over all possible 
phases $\Omega$ which appear in the change of the mixed Chern-Simons term 
(\ref{mixed}) 
arising in the effective theory of the putative superconducting 
phase 
after integrating out massive 
(fermionic (holon) and bosonic(spinon) ) degrees 
of freedom~\cite{dorey} (c.f. discussion in previous subsection).
If we start from the classical theory, then as discussed above, this
would imply quantization of the electromagnetic flux which would lead
to pairing, $\Phi = m\pi/2e$, $m$ an integer. 
When quantum fluctuations of the monopole are added
then one has to consider the path integral over the phases $\Omega$:
\begin{equation} 
\int _0^{2\pi} d\Omega e^{im\Omega} =\delta_{m0} 
\label{washout}
\end{equation} 
which implies that any quantization of the electric flux is washed out,
hence no flux quantization. This implies that the superconducting 
phase would correspond only to a phase where the monopoles are 
{\it bound in pairs 
with their antimonopoles}~\cite{mavromatos}, which is in agreement
with the Kosterlitz-Thouless nature of the superconductivity 
of the present composite model, as 
well as that of \cite{dorey}.

\begin{figure}[t]
\epsfxsize=3.5in
\bigskip
\centerline{\epsffile{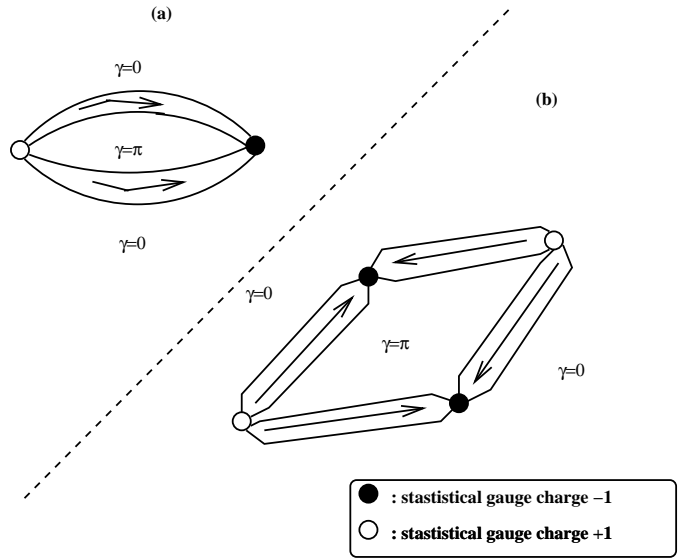}}
\caption{Examples of domain wall structures for $p=2$ in compact 
N=2 supersymmetric Abelian gauge theories:
(a) a configuration in $p=2$ SQED with massive matter, 
with a pair of heavy monopole charges, of statistical 
gauge charge +1 (dark blob) and -1 (white blob),
(b) a configuration in pure U(1) N=2 supersymmetric gauge theory
with two pairs of heavy monopole charges, of statistical 
gauge charge +1 and -1. 
The domain wall
structures
confine in their interior 
excitations of the statistical `electric' field (arrows), and separate
the two vacua of the theory labeled by the dual photon v.e.v. $\gamma=0$ and 
$\gamma=\pi$. In the actual condensed matter situation, with the 
statistical gauge field
being given by (\ref{pviol}), there is a real electric current flow in the 
domain wall structures, which prompts one to identified them with 
the 
stripes  observed in the stripe phase of the underdoped cuprates. 
\label{fig:walls}}
\end{figure}

The stable monopole phase, however, 
has other attractive properties which are tempting us to identify 
with a non-superconducting stripe phase, and which 
we now proceed to discuss briefly. We shall restrict ourselves
below to the effect of supersymmetric monopoles. For 
a possible connection of monopole and stripe phases in 
non-supersymmetric composite theories the reader is 
advised to see the analysis in
ref.~\cite{sarbdual}. 
The gauge monopole phase 
is characterised by the existence of 
domain walls of a given flux,
which, as discussed in \cite{strassler, intril} and reviewed briefly 
in the Appendix, form groups of, say $p$ of them,
separating $p$ different vacua of the theory, emanating from 
a heavy particle/monopole configuration 
of positive statistical charge, and/or ending in one with negative charge
(see figure \ref{fig:walls}).  For the record we mention that 
exact results as far as the structure of the domain walls is concerned
have been derived only in the case where one assumes a {\it bare mass} $m_b$
for the matter superfields, which contributes to a 
a bare superpotential term of the form $m_b Q_1 Q_2$. 
In such a case, the moduli space of SQED turns out to be qualitatively
similar to that of pure compact $U(1)$ gauge theory, given that one may 
integrate out in a Wilsonian effective action the massive degrees of freedom,
to obtain the effective theory of the massless one, whose spectrum
is similar to pure U(1) in the Coulomb phase. The superpotential, as mentioned
in  the Appendix, is protected by non-renormalization theorems~\cite{intril},
and  hence the above result turns out to be exact. 

In conformal SQED, however, which is 
characterised by the absence of a mass term for 
the matter multiplets, such a mass term is generated 
dynamically in the Coulomb phase of the theory, but such a mass term
arises from the kinetic terms of the supermultiplet, and thus 
does not appear in the superpotential. In that case, one also expects 
the dynamical appearance of domain walls of similar nature (qualitatively)
to that in figure \ref{fig:walls}, but unfortunately there are no exact 
results available to this case. In our composite theory a bare mass term
for the {\it composite} matter multiplets is not in contradiction with the fact that
we work here with nodal excitations. A composite mass term corresponds to 
contact quartic (or higher) interactions among the fundamental constituents,
(spinon $z$ and holons $\psi_i$), and as such one may have 
it in the effective 
theory. To turn the logic around, even if such terms are not 
appearing in the microscopic theory, one may add them by hand in order 
to guarantee the exactness of the result, and then turn them off adiabatically
claiming that qualitatively the information about the domain wall
phase does not change.

Along each domain wall,
separating two different vacua, there is an excitation of the flux 
of the gauge field. From a point of view of a phase (moduli space) 
diagram, the Coulomb branch of compact SQED is split into two, 
in one of which there is a stable monopole plasma phase, as discussed
in the appendix, where such stripe phenomena occur. The presence of 
monopoles are responsible for giving a small but finite mass to the 
statistical gauge field $A_\mu^4$, and hence the superconductivity,
in the sense of the satisfaction of the Landau criterion 
of figure \ref{fig:sc}, is destroyed. This phase is therefore also a 
pseudogap phase, but there are no bosonic pair excitations as in the Higgs 
phase. There is also pairing, which is obtained by means of 
the mixed Chern-Simons terms in the effective lagrangian, upon coupling
the theory to an external electromagnetic field. In \cite{strassler}
it was argued that the domain walls  
in the configuration of figure \ref{fig:walls}(a) attract each other,
which may result in a bigger domain wall with twice the flux.

{}From our composite construction the gauge field 
(\ref{pviol})
carries 
holon current excitations, and hence the interior of the stripes 
is characterised by real electric charge flow. In view of this 
one might be tempted to identify 
this monopole pseudogap phase with the stripe phase, in analogy 
with the non supersymmetric case of ref.~\cite{sarbdual}.
The important advantage of the current model is that it is 
Abelian, and thus the results may provide exact (at least qualitatively)
non perturbative information,
in contrast to the non-Abelian non-supersymmetric 
model of \cite{sarbdual}, where one 
relies on perturbative arguments of weakly coupled gauge theories
(another difference of the current work is that here the gauge fields
are induced from fundamental constituents, and one 
has a detailed knowledge of their microscopic 
constituent structure in terms of spinon and holons, and hence one
does not have
to rely on mean-field arguments on their form, based upon phase
fluctuations in the spin-charge separation ansatz, which was the 
case of the various constituent approaches so far).

\paragraph{} 

\underline{N=2 SU(2)-like supersymmetry and superconductivity}

\paragraph{} 
Before closing we would like to make an important 
comment regarding the 
absence of the stable monopole plasma phase in N=2 supersymmetric 
SU(2) non-Abelian gauge model of \cite{affleck}. 
As emphasized in \cite{strassler}, and reviewed in the 
Appendix, the SU(2) case is characterised only by monopoles of charge +1,
and antimonopoles of charge -1, unlike SQED where both charges are present.
In such non-Abelian theories, therefore, the stripe (domain wall) 
phase, due to monopole plasma, would be 
absent. On the other hand, as a result of the existence of still 
non-trivial configurations with charge +1, one might still consider
large gauge temporal transformations in the respective effective action 
which would lead to pairing as in (\ref{fluxq}). Thus the existence
of such N=2 SU(2)-type supersymmetric points would be equivalent to 
the onset of superconductivity in the model. 

From a microscopic point of view 
one might think of the possibility of approaching such a N=2 supersymmetric
SU(2) point by letting the fugacities ${\tilde h}$ 
of the -1 charge monopole excitations, appearing in the 
superpotential of the theory (c.f. Appendix for notation),  
depend on the doping concentration in the sample as: 
${\tilde h} \sim (\delta - \delta_c)^\gamma$, where $\delta$ is the 
doping concentration, $\gamma > 0$ is some critical 
exponent,  
and $\delta_c$ is a critical value marking the 
on set of such `N=2 SU(2) supersymmetry-like' regimes. 

This could be a viable way of entering from a pseudogap (stripe) 
phase to 
superconductivity, which was the original way envisaged in 
\cite{farakos,mav}, and we still think
describes (in some sense) the situation encountered in nature. 
In other words, the appearance of a N=2 SU(2)-like
supersymmetric 
situation (corresponding to the vanishing of the fugacity of the charge +1
monopoles as one varies the doping concentration in the sample) 
would destabilize the pseudogap stripe/monopole plasma phase in favour
of the Kosteritz-Thouless type superconductivity scenario of \cite{dorey}, reviewed above.

Such a possibility should be explored further
in terms of microscopic and/or composite models. At present it is 
not possible to construct explicitly 
a fully non-Abelian SU(2) gauge model of composite
supersymmetric excitations, for reasons explained in the text above.
Thus, at least at present, 
the N=2 ``SU(2)-like'' points have to be viewed only as 
points of SQED compact theories with zero fugacities for monopoles
of charge -1, characterised by the absence of a stable monopole
phase of domain wall type. 
However, the construction of a fully non-Abelian 
SU(2) composite theory may be a possibility, and it certainly constitutes
an issue worthy of further exploration.

\paragraph{} 
\underline{A Non-trivial Infrared Fixed Point and Non-fermi liquid Behaviour}

\paragraph{} 
A final comment we would like to make, concerning exact results in 
N=2 supersymmetric (2+1)-dimensional theories, is associated
with the existence of a non-trivial (non Gaussian) {\it infrared fixed point}.
In N=2 theories~\cite{intril,strassler} such a statement becomes qualitatively
{\it exact} by the discovery of appropriate {\it dual} models, with which
SQED lies in the {\it same renormalization group universality class} 
in the {\it infrared}. This dual model
is the so-called XYZ supersymmetric model, with a cubic (in superfields 
X,Y,Z) superpotential, mentioned briefly in the appendix. 

Such a property implies that the nodal supersymmetric liquid exhibits
a non-fermi liquid behaviour, if one defines the deviation from the 
fermi liquid behaviour as being essentially equivalent to the 
absence of a Landau (Gaussian) fixed point at low energies. 
Arguments why this is so are given in \cite{nflb}. 

Our N=2 supersymmetric model is similar in this respect to N=1 or N=0
(non supersymmetric) models of three dimensional Abelian gauge theories,
for which it has been argued that there is a non-trivial 
infrared structure~\cite{aitchliq,papavas}. 
The important difference on the N=2 case is, however,
that the result is {\it exact}, in contrast to these other cases,
where the result was argued on the basis of approximate 
resummation Schwinger-Dyson techniques, such as large N-flavour number
of fermions~\cite{aitchliq}, 
or the so-called pinched technique using appropriately
resummed amputated fermion-gauge-boson vertices~\cite{papavas}.

\section{Discussion and Outlook}

In this article 
we have discussed a way of obtaining (in the continuum limit) 
a N=2 supersymmetric effective lagrangian, describing the dynamics
of $d$-wave 
nodal composite excitations 
made out of constituent spinons and holons within 
a
spin-charge separation framework 
in a certain regime of the parameter space
of extended $t-j$ models of 
doped antiferromagnets~\cite{mav}. We have paid particular attention
to discussing how the continuum limit of the constituent 
microscopic theory is taken, and how, 
once this is done, the composite supersymmetric structure emerges.
Of particular importance was the existence of an {\it even number}
of pairs of fermi-surface-$d$-wave nodes, which implies 
the emergence of 
an {\it even} number of composite chiral matter supermultiplets $Q_{1,2}$ 
in the effective (continuum)  
nodal composite lagrangian, 
and thus the possibility of {\it parity} conserving
effective theories. 
The (statistical) 
gauge composite supermultiplet, on the other hand, whose existence
was necessitated by supersymmetry~\cite{compos,mav}
expresses frustration of (interaction 
between) those pairs of nodes. The nodal pairs couple with 
opposite statistical gauge charge for energetic 
reasons (parity conservation).

Although our composite construction 
has taken place in the continuum limit
of the constituent theory, nevertheless
our hope is that this
supersymmetric continuum theory belongs to the same {\it universality class}
as the corresponding composite lattice model. In other words,
the two theories may differ only by a number of 
irrelevant operators in a renormalization group sense. 
This, however, remains to be demonstrated, and we hope to come back 
to such issues in future works. 

Basing our considerations on this naive continuum limit we have demonstrated
the existence of several {\it exact results} concerning the phase structure
of the composite continuous theory, including a passage from the 
pseudogap to an unconventional superconducting pairing state 
in the spirit of ~\cite{dorey}, as well as the existence of a stripe phase,
and the non-fermi liquid behaviour of the relativistic nodal composite 
liquid. As regards the latter property, it should be mentioned 
that this was the result of 
the existence of a non-trivial infrared (low energy) fixed point in the 
renormalization group analysis of (2+1)-dimensional 
N=2 supersymmetric gauge 
theories under consideration~\cite{intril,deboer}. 

As a possible outlook of the current approach we remark on the possibility 
of having {\it Galilean supersymmetry} for the theory away from 
the nodes, in the same regime of the parameter space of the microscopic model
which leads to the Lorentzian nodal supersymmetry. Works with Galilean
supersymmetry  
do exist in the field theoretic literature in (2+1)-dimensions~\cite{lozano}, 
but at present we are far from translating such field-theoretic
results to excitations pertaining to realistic
condensed-matter situations, describing the physics  
away from nodes in the fermi surface
of $d$-wave antiferromagnetic superconductors. We intend to embark 
on a study of such 
important issues in the near future.

We do believe that the present work, together with our previous works
on this topic~\cite{mav,compos,ams,sarbdual}, opens up the way for a 
possible formal
understanding of the phase diagram (at least at zero or very low temperatures)
of doped antiferromagnets, and hence high temperature superconductors, 
in an analytically {\it exact} way, at least in a specific regime 
of the parameters of the relevant microscopic lattice models, 
where extended N=2 supersymmetries between spinon and holon
degres of freedom in a spin-charge
separation framework do exist. In this way one may hope to  
extrapolate such exact results {\it in a qualitative manner} 
away from such parameter
regions, where the supersymmetries are explicitly broken. 
At present we do not know precisely how to do such an extrapolation,
but we are strongly encouraged by the current results so as to continue
pursuing this research further.

\section*{Acknowledgements} 

We acknowledge informative discussions with G. Lozano.
This work is supported by the Leverhulme Trust (U.K.).

\section*{Appendix}

\subsection*{N=2 Supersymmetric Quantum Electrodynamics (SQED): 
A review of the formalism} 

A $N=2$ supersymmetric theory in 2+1 dimensions can be obtained by dimensional reduction of
a $N=1$ supersymmetric theory in 3+1 dimensions, in which the two two-component Grassmann 
coordinates of superspace lead to one complex two-component Grassmann coordinate $\theta$.
We have then the following properties:

\bea
&&\ol\theta\theta=(\ol\theta\theta)^\star\nonu
&&\ol\theta\gamma^\mu\theta=(\ol\theta\gamma^\mu\theta)^\star.
\eea

\nin The Abelian vector 
supermultiplet contains the following degrees of freedom: a gauge field $A_\mu$, 
a complex gaugino $\lambda$ and a real scalar $\rho$. The (real) superfield $V$ 
corresponding to the vector supermultiplet contains in addition
an auxiliary field $D$, which is essential for the generation of a scalar
potential, as will be seen later on. 
$V$ has the following expansion in the Grassmann variables $\theta$
and $\ol\theta$:

\be
V=\ol\theta\theta\rho+(\ol\theta\gamma^\mu\theta)A_\mu+i\ol\theta\theta^\star(\ol\theta^\star\lambda)
-i\ol\theta^\star\theta(\ol\theta\lambda^\star)+\frac{1}{2}|\ol\theta\theta|^2D.
\ee

\nin The gauge kinetic term is given by the square of the linear superfield $\Sigma=\ol D D V$ 
\cite{strassler}, where the supercovariant derivative is \cite{rocek}

\be
D^\alpha=\frac{\partial}{\partial\theta^\star_\alpha}+i(\br\partial\theta)^\alpha.\nonu
\ee

\nin We have then for the gauge multiplet kinetic term

\bea
{\cal L}_{gauge}&=&\int d^2\theta d^2\ol\theta ~ \frac{1}{4}\Sigma^2\nonu
&=&\frac{1}{4g^2}\left(-F_{\mu\nu}F^{\mu\nu}-2\rho\Box\rho+i\left(\ol\lambda\br\partial\lambda
+(\ol\lambda\br\partial\lambda)^\star\right)+2D^2\right)\nonu
&=&-\frac{1}{4g^2}F_{\mu\nu}F^{\mu\nu}+\frac{1}{2g^2}\partial_\mu\rho\partial^\mu\rho+
\frac{i}{g^2}\ol\lambda\br\partial\lambda+\frac{1}{2g^2}D^2+\mbox{surface terms},
\eea

\nin where the surface terms do not contribute after integration over space-time.

We describe matter with a chiral superfield $Q$ which 
contains a scalar $\Phi$, its superpartner $\Psi$ and an auxiliary field $F$, 
all complex, and its expansion in $\theta$ and $\ol\theta$ is

\be
Q=\Phi+\ol\theta^\star\Psi+\hf\theta^2F+\frac{i}{2}(\ol\theta\gamma^\mu\theta)\partial_\mu\Phi
+\frac{i}{2}\ol\theta^\star\ol\theta\theta\br\partial\Psi
-\frac{1}{8}|\ol\theta\theta|^2\Box\Phi.
\ee

The matter kinetic term including its coupling to the vector multiplet is given by

\bea
{\cal L}_{matter}^{kin.}
&=&\int d^2\theta d^2\ol\theta ~ Qe^{2gV} Q^\star\\
&=&\frac{1}{4}\partial_\mu\Phi\partial^\mu\Phi^\star
-\frac{1}{8}(\Phi\Box\Phi^\star+\Phi^\star\Box\Phi)
+\frac{i}{4}(\ol\Psi\br\partial\Psi+(\ol\Psi\br\partial\Psi)^\star)\nonu
&&+\frac{i}{2}g(\Phi^\star A_\mu\partial^\mu\Phi+\Phi A_\mu\partial^\mu\Phi^\star)+
\frac{g^2}{2}A_\mu A^\mu\Phi\Phi^\star-g\ol\Psi\br A\Psi\nonu
&&+g\frac{i}{2}(\Phi^\star\ol\lambda^\star\Psi-\Phi\ol\lambda\Psi)
-g^2\rho^2\Phi\Phi^\star-g\rho\ol\Psi\Psi+\hf F F^\star+\frac{g}{2}D\Phi\Phi^\star\nonu
&=&\hf D_\mu\phi\left(D^\mu\Phi\right)^\star
+\frac{i}{2}\ol\Psi\br D\Psi+\frac{i}{2}g(\Phi^\star\ol\lambda^\star\Psi-
\Phi\ol\lambda\Psi^\star)\nonu
&&-\frac{g^2}{2}\rho^2\Phi\Phi^\star-\frac{g}{2}\rho\ol\Psi\Psi+\hf F F^\star
+\frac{g}{2}D\Phi\Phi^\star+\mbox{surface terms},\nonumber
\eea

\nin where $g$ is the gauge coupling and $D_\mu=\partial_\mu+igA_\mu$.

To ${\cal L}_{matter}^{kin.}$ can be added a mass term, or more generally a self-interaction term via the 
superpotential $W(Q)$ which contributes to the Lagrangian as follows

\bea
{\cal L}_{matter}^{self.}
&=&\int d^2\theta  W(Q)~+~\mbox{h.c.}\\
&=&\frac{\partial W}{\partial Q} F-\frac{\partial^2W}{\partial Q^2}\ol\Psi\Psi~+~\mbox{h.c.},\nonumber
\eea

\nin where the derivatives of the superpotential are to be taken at $\theta=\ol\theta=0$.
A mass term for the matter can be taken into account with the superpotential $W(Q)=mQ^2$.

Finally, for an Abelian theory, a Fayet-Iliopoulos term can be added:

\be
{\cal L}_{F.I.}=\int d^2\theta d^2\ol\theta ~ \left(-2g\phi_0^2V\right)=-g\phi_0^2 D.
\ee

We consider 2 matter flavors $Q_1$ and $Q_2$, with opposite couplings
so as to avoid the parity anomalies
and the most general bare Lagrangian reads, when we disregard the surface terms

\bea\label{superL}
{\cal L}&=&\int d^2\theta d^2\ol\theta ~\left(\frac{1}{4}\Sigma^2+Q_1e^{2gV}Q_1^\star+Q_2e^{-2gV}Q_2^\star
-2g\phi_0^2V\right)\\
&&~~~~~~+\int d^2\theta W(Q_1,Q_2)~+~\mbox{h.c.}\nonu
&=&-\frac{1}{4}F_{\mu\nu}F^{\mu\nu}+\frac{1}{2}\partial_\mu\rho\partial^\mu\rho
+\frac{i}{2}\ol\lambda\br\partial\lambda\nonu
&&+\hf D_\mu\Phi_1\left(D^\mu\Phi_1\right)^\star+\frac{i}{2}\ol\Psi_1\br D\Psi_1
-\frac{g^2}{2}\rho^2\Phi_1\Phi_1^\star
-\frac{g}{2}\rho\ol\Psi_1\Psi_1\nonu
&&+\hf D_\mu\Phi_2\left(D^\mu\Phi_2\right)^\star+\frac{i}{2}\ol\Psi_2\br D\Psi_2
-\frac{g^2}{2}\rho^2\Phi_2\Phi_2^\star
+\frac{g}{2}\rho\ol\Psi_2\Psi_2\nonu
&&+g\frac{i}{2}(\Phi_1^\star\ol\lambda^\star\Psi_1-\Phi_2^\star\ol\lambda^\star\Psi_2-\mbox{c.c.})\nonu
&&+\frac{1}{2}D^2+\frac{g}{2}D(|\Phi_1|^2-|\Phi_2|^2)-\frac{g}{2}\phi_0^2D
+\hf |F_1|^2+\hf |F_2|^2\nonu
&&+\left(\frac{\partial W}{\partial Q_1} F_1+\frac{\partial W}{\partial Q_2} F_2
-\frac{\partial^2W}{\partial Q_k\partial Q_l}\ol\Psi_k^\star\Psi_l~+~\mbox{h.c.}\right)\nonumber
\eea

\nin where $k,l=1,2$ and

\bea 
D_\mu(...)_1&=&\partial_\mu(...)_1+igA_\mu(...)_1\nonu
D_\mu(...)_2&=&\partial_\mu(...)_2-igA_\mu(...)_2.
\eea

The scalar potential is obtained when writing 
the equations of motion of the auxiliary fields, which are

\bea
&&F_1=-2\left(\frac{\partial W}{\partial Q_1}\right)^\star,
~~~~~F_2=-2\left(\frac{\partial W}{\partial Q_2}\right)^\star,\\
&&D=-\frac{g}{2}(|\Phi_1|^2-|\Phi_2|^2-\phi_0^2)\nonumber,
\eea

\nin and lead to the following potential

\bea\label{pot}
U&=&\frac{g^2}{2}\rho^2(|\Phi_1|^2+|\Phi_2|^2)-\hf|F_1|^2-\hf|F_2|^2\nonu
&&-\hf D^2-\frac{g}{2}D(|\Phi_1|^2-|\Phi_2|^2)+\frac{g}{2}\phi_0^2D\nonu
&&-\left(\frac{\partial W}{\partial Q_1} F_1+\frac{\partial W}{\partial Q_2} F_2~+~\mbox{h.c.}\right)\nonu
&=&\frac{g^2}{2}\rho^2(|\Phi_1|^2+|\Phi_2|^2)
+\frac{g^2}{8}(|\Phi_1|^2-|\Phi_2|^2-\phi_0^2)^2\nonu
&&+2\left(\left|\frac{\partial W}{\partial Q_1}\right|^2+\left|\frac{\partial W}{\partial Q_2}\right|^2
~+~\mbox{h.c.}\right).
\eea

\nin The on-shell Lagrangian is finally

\bea\label{onshelllagrang}
{\cal L}_{on-shell}&=&-\frac{1}{4}F_{\mu\nu}F^{\mu\nu}+\frac{1}{2}\partial_\mu\rho\partial^\mu\rho+
\frac{i}{2}\ol\lambda\br\partial\lambda\\
&&+\hf D_\mu\Phi_1\left(D^\mu\Phi_1\right)^\star+
\hf D_\mu\Phi_2\left(D^\mu\Phi_2\right)^\star
+\frac{i}{2}\ol\Psi_1\br D\Psi_1+\frac{i}{2}\ol\Psi_2\br D\Psi_2\nonu 
&&+i\frac{g}{2}(\Phi_1^\star\ol\lambda^\star\Psi_1-\Phi_2^\star\ol\lambda^\star\Psi_2-\mbox{c.c.})
-\frac{g}{2}\rho(\ol\Psi_1\Psi_1-\ol\Psi_2\Psi_2)-U(\rho,\Phi_1,\Phi_2),\nonumber
\eea

\nin where $U$ is given by Eq.(\ref{pot}). Note that the Higgs self-coupling is $g^2/8$, as found
in \cite{edelstein,alex} due to 
the elevation of a $N=1$ supersymmetry to $N=2$.

\subsection*{Phases of N=2 SQED}

Let us discuss the phases of the theory. 
The analysis is outlined in \cite{strassler} and in more details in 
\cite{intril}. Here we review only the basic properties to be 
used in our physical discussion with respect to the various phases
of the nodal supersymmetric liquid. 
The basic toll for understanding the various phases of the theory 
is the so-called moduli space of supersymmetric vacua.

To understand the topology of the moduli space in the SQED case
we first note that 
classically N=2 SQED without matter 
has a phase of vacua labeled by the vacuum expectation value of the 
scalar field $\rho$ in the N=2 vector multiplet $(\rho, A_\mu, \lambda) $,
where $\lambda $ is the photino, $A_\mu$ the photon, and $\rho$ the scalar
(in what follows we shall refer
to the gauge boson of SQED as a `photon' although from our physical point 
of view this will be the statistical photon, not to be confused
with the carrier of the real electromagnetic interactions).

Quantum mechanically, one can replace the gauge field by its dual,
$\varepsilon_{\mu\nu\rho}\partial^\nu A^\rho = \partial_\mu \gamma$,
where $\gamma$ is a scalar periodic under $\gamma \to \gamma + 2\pi$.
One defines then a chiral superfield $T$ whose lowest component
is $\rho + i\gamma $. The moduli space is then 
given by a `metric' space whose coordinates are given by the v.e.v.s 
of the scalar fields generating a supersymmetric vacuum 
($\rho$ and $\gamma$ in our case).
As discussed in \cite{seiberg},
the precise geometry of this space at a quantum level determines
the phase space structure of the supersymmetric theory. 
In the SQED case, 
the Coulomb branch of the moduli space 
is 
classically the cylinder defined by $<T>$.
Due to the periodicity of the dual field $\gamma$, $T$ is a constrained 
chiral 
superfield and the actual good single-valued (and unconstrained)
superfields for the description of the moduli space are $e^{\pm T}$ such that the Coulomb 
branch is described by $<e^{\pm T}>$.

In a non compact SQED there is no monopole phase.
Things get more complicated in case of compact SQED, which is the case
in which the compact Abelian gauge field may also be embedded in a
non-Abelian subgroup. We shall stress the important differences
between the two cases later on.

\begin{figure}[t]
\epsfxsize=3in
\bigskip
\centerline{\epsffile{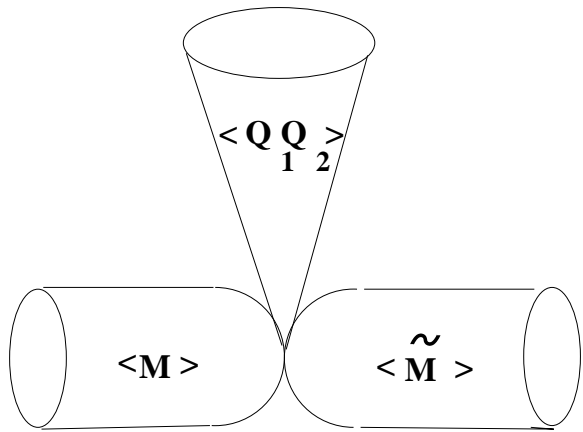}}
\caption{The quantum moduli space of N=2 SQED with conformal 
matter and zero superpotential. Its topology 
is an exact result. The phase where $<Q_1 Q_2> \ne 0$ 
is the Higgs phase, while the two branches characterised by $<M=e^T> \ne 0$ 
and $<{\tilde M}=e^{-T}> \ne 0$, with $T$ the dual of the vector 
superfield, constitute the Coulomb phase.  In case there is a bare 
mass for chiral multiplets, corresponding to a superpotential term 
$mQ_1 Q_2$, the conformal point at the origin is replaced by
a small neck, of finte thickness. 
\label{fig:moduli}}
\end{figure}

At the moment we note that,
since a supersymmetric theory has zero vacuum energy,
the scalar field vacuum expectation values (v.e.v.) 
$<\rho>,<\Phi_1>,<\Phi_2>$ must satisfy:
\be
U(<\rho>,<\Phi_1>,<\Phi_2>)=0.
\ee
\nin We will assume that there is no bare superpotential $W=0$. 
This is consistent with our microscopic condensed-matter applications
we are interested in; such a superpotential is generated
by non perturbative effects in the physically relevant case of 
compact SQED, 
as we shall discuss later on. 
Then, for a
supersymmetric vacuum to occur we have the following 
possibilities: 

\begin{itemize}
\item In the presence of the Fayet-Iliopoulos term, 
the only possibility is a Higgs phase where 
$<|\Phi_1|^2>=<|\Phi_2|^2>+\phi_0^2\ne 0$ and $<\rho>=0$. 
In this phase, the matter fields remain
massless and the gauge field acquires the mass $<|\Phi_1|^2>+<|\Phi_2|^2>$. 
The theory is then the supersymmetric Abelian Higgs model \cite{edelstein}.

\item Without Fayet-Iliopoulos term ($\phi_0=0$),
which is the case relevant for our 
condensed-matter application here, there is also, 
besides the Higgs phase where
$<\Phi_1>=<\Phi_2>\ne 0$ and $<\rho>=0$, a Coulomb phase where $<\Phi_1>=<\Phi_2>=0$ and $<\rho>\ne 0$. 
In this phase,
the matter fields acquire a mass given by $<\rho>$ and the gauge field remains massless. 
In the moduli space, the Higgs branch is
a two dimensional conical (real) surface and has its summit at the origin $<\Phi_1>=<\Phi_2>=<\rho>=0$
\cite{intril,strassler}.
The Higgs and Coulomb branches intersect at the origin of the moduli space. 
\end{itemize}

In the Lagrangian (\ref{superL}), only the superpotential $W$ 
is protected by non-renormalization theorems 
and thus cannot be generated by perturbative quantum corrections if it is not present at the tree level. However,
the Higgs and Coulomb phases do get perturbative quantum corrections. 
But, as explained in \cite{intril}, the topology of the Higgs phase
is not changed and only the Coulomb 
branch changes qualitatively as it is shown 
in figure \ref{fig:moduli}.
Another possibility is to have non-perturbative quantum correction, coming from
solitons if the Abelian gauge field is compact. 
Such corrections can generate a superpotential
for the chiral field dual to the vector field, 
as will be discussed next.

\paragraph{}  
\underline{Dual of N=2 SQED} 

As an important remark, before we proceed onto a discussion
of the compact case, we would like to mention that
one can find~\cite{strassler}  
an {\it exact dual}, in the sense of 
identical moduli space and  
spectrum of gauge invariant operators (i.e., they belong to
the same {\it universality class}, 
of the N=2 SQED, which 
is described by the so-called XYZ supersymmetric model.
This model consists of three chiral superfields $X,Y,Z$ with 
a superpotential ${\cal W}=eXYZ$, 
where $e$ is a (dimensionful) coupling constant. The compact case then, and the associated 
effects of the monopole/instantons
on the dynamics of SQED can be understood by studying various 
superpotential configurations of the dual model, and this is the 
approach followed in \cite{strassler}, whose results, as far as SQED
monopole phase is concerned, are reviewed
below. It is this exact duality that can be used 
in extracting exact information on the non-trivial 
infrared fixed point structure of SQED, with consequences
on the non-fermi liquid behaviour of the nodal systems, as we 
discussed in the main text.

\paragraph{}  
\underline{Compact SQED: the important physical differences} 

We now make some comments for compact SQED. This 
will help the reader understand the 
important physical 
differences from the non compact case. 
In compact SQED $M=e^T,~{\tilde M}=e^{-T}$
are chiral operators representing point-like
instantons (Dirac magnetic monopoles in three dimensions), 
with charges 1, -1 respectively.  
The complex conjugates $M^\dagger,~{\tilde M}^\dagger$ have charges 
-1,1 respectively. 

There is logarithmic confinement in SQED due to its low dimensionality,
but also there is linear confinement in the phase where there is a monopole
plasma. There is a detailed discussion in \cite{strassler}, to which we refer
the interested reader. The proof that the monopole plasma phase 
leads to linear confinement is similar to that of \cite{polyakov}.

Formally, the presence of a linearly confined phase is described 
by adding to the SQED Lagrangian a superpotential term (interactions)
for the chiral operators $M,~{\tilde M}$. 
Consider $p$ a positive integer, and the superpotential 

\begin{equation}\label{superpot} 
{\cal W}(T) = hM^{p/2} + \tilde h\tilde M^{p/2} 
\end{equation} 

\nin where $h$, (${\tilde h}$) denote densities (fugacities) 
of monopoles with charges 1, (-1) respectively.
${\cal W}(T)$ is generated by instanton effects and 
in general $h \ne {\tilde h}$. There could be various 
other terms in the superpotential. Most of the results
reviewed below for the monopole phase of SQED are obtained
by virtue of the dual model~\cite{strassler}. 

If $h={\tilde h}$ then there is a symmetry $T \to -T$ 
in the theory, and in such a case there are equal  
amounts of monopoles of charge $\pm 1$ and antimonopoles
of charge $\mp 1$, implying vacua symmetric about 
$\rho =0$. In the general case $h \ne {\tilde h}$ 
one may reduce the effect of the charge -1 monopoles
${\tilde M}$ relative to the charge 1 monopoles $M$
by a field rescaling $M \to A~M$, ${\tilde M} \to {\tilde M}/A$,
where $A=\left(\frac{{\tilde h}}{h}\right)^{\frac{1}{p}}$.
This is equivalent to 
{\it shifting} the symmetry to 
$ T \to -T + {\rm ln}\left(\frac{{\tilde h}}{h}\right)^{\frac{2}{p}} $,
which essentially acts as a shift of $\rho$. 

In the limit ${\tilde h} \to 0$ stable supersymmetric vacua 
occur therefore only for $<\rho> \to \infty$, and hence in that
special case there is no stable supersymmetric vacuum with 
monopole plasma and  
linearly confined phase for finite vevs of $\rho$.
 
This last case, for $p=2$, is the case of the non abelian SU(2) Georgi-Glashow 
N=2 supersymmetric model discussed in \cite{affleck}.
Indeed an index theorem~\cite{strassler,deboer} 
guarantees that the SU(2) case has only monopoles 
of charge +1 and antimonopoles of charge -1. In this case ${\tilde h}=0$ 
and, hence, in view of the above discussion there is no monopole plasma 
phase. For a detailed discussion on SU(2) with chiral matter
multiplets see section 3.2 of ref. \cite{deboer}. 

For a generic exponent $p$ in (\ref{superpot}), which, as discussed in 
\cite{strassler},  
must be integer to ensure vanishing vacuum energy,
as required by supersymmetric vacua, there are $p$ such vacua
separated by $p$ domain walls (flux strings) which meet at a vortex of 
SQED (which exist because of the complex scalar fields of the gauge or chiral 
multiplets). For even $p$ the strings 
may connect pairs of vortices-antivortices. 
We remind the reader that the embedding in the SU(2) occurs only for $p=2$.
On each flux string the gauge field dual is excited 
in the sense of the string carrying non trivial flux.

On the other hand, in the case where ${\tilde h} \to 0$ there is no
stable monopole plasma phase and one has the superconducting regime
according to the exact masslessness of the statistical photon.
One may then have an elevation to the SU(2) case at such points,
given that in SU(2) N=2 supersymmetric theories there are only monopoles
of charge +1, and antimonopoles of charge -1, unlike SQED where both
kinds of charges are present. The absence of a stable monopole phase
in SU(2) N=2 supersymmetric 
gauge theories is confirmed 
by independent arguments~\cite{deboer}, based on the 
so-called Wilsonian effective action to gauge theories, where 
massive degrees of freedom are being integrated out.

\end{document}